\documentclass[12pt]{article}
\usepackage{amsmath}
\usepackage{amssymb}
\begin{document}
\begin{center}
{\bf {\Large   {(Anti-)BRST and (Anti-)co-BRST Symmetries  in 2D non-Abelian Gauge Theory: Some Novel Observations}}}

\vskip 3.0cm

{\sf S. Kumar$^{(a)}$, B. K. Kureel$^{(a)}$, R. P. Malik$^{(a,b)}$}\\
$^{(a)}$ {\it Physics Department, Institute of Science,}\\
{\it Banaras Hindu University, Varanasi - 221 005, (U.P.), India}\\

\vskip 0.1cm

\vskip 0.1cm

$^{(b)}$ {\it DST Centre for Interdisciplinary Mathematical Sciences,}\\
{\it Institute of Science, Banaras Hindu University, Varanasi - 221 005, India}\\
{\small {\sf {e-mails: sunil.bhu93@gmail.com; bkishore.bhu@gmail.com;  rpmalik1995@gmail.com}}}

\end{center}

\vskip 2cm

\noindent
{\bf Abstract:} We discuss  the nilpotent  Becchi-Rouet-Stora-Tyutin (BRST), anti-BRST and 
(anti-)co-BRST symmetry transformations and derive their corresponding conserved 
charges in the case of a two (1+1)-dimensional (2D) {\it self-interacting} non-Abelian gauge theory (without any interaction with  
matter fields). We point out a set of  {\it novel} features that emerge out in the
BRST and co-BRST analysis of the above 2D gauge theory. The algebraic structures of the symmetry operators (and corresponding 
conserved charges) and their relationship with the cohomological operators of differential
 geometry are established, too. To be more precise, we demonstrate the existence of a 
{\it single}\,Lagrangian  density that respects the continuous symmetries which obey 
proper algebraic structure  of the cohomological operators of differential geometry. 
In literature, such observations have been made for the {\it coupled}  (but equivalent) Lagrangian densities 
of the 4D non-Abelian gauge theory.
We lay emphasis on the existence and properties of the Curci-Ferrari (CF) type restrictions 
in the context of (anti-)BRST and (anti-)co-BRST symmetry transformations and pinpoint 
their key differences and similarities. All the observations, connected with the (anti-)co-BRST 
symmetries, are {\it completely} novel.

\vskip 2cm



\newpage
\section{Introduction}

The  principles of {\it local} gauge theories are at the heart of a precise theoretical description of electromagnetic, weak and strong 
interactions  of nature. One of the most intuitive, geometrically rich and mathematically elegant  methods to quantize such
 kind of theories is the Becchi-Rouet-Stora-Tyutin (BRST) formalism [1-4]. In this formalism, the {\it local} gauge symmetries of the 
{\it classical} theories are traded with the (anti-)BRST symmetries at the {\it quantum} level where unitarity  is satisfied at any
 arbitrary order of perturbative computations. These (anti-)BRST symmetries are fermionic (i.e. supersymmetric-type) in nature and, 
therefore, they are nilpotent of order two. However, these symmetries absolutely {\it anticommute}
 with each other. This latter property encodes the
linear independence of these symmetries. Hence, the BRST and anti-BRST symmetries have their own identities.

The (anti-)BRST symmetry transformations are  {\it fermionic} in nature because they transform a bosonic field into its fermionic
counterpart and {\it vic{\`e}-versa}. This is what precisely happens with the supersymmetric (SUSY) transformations which are {\it also}
fermionic in nature. However, there is a decisive difference between the {\it two}. Whereas the (anti-)BRST symmetry transformations are absolutely 
anticommuting in nature, the anticommutator of {\it two} distinct SUSY transformations always produces the spacetime translation 
of the field on which it (i.e. the anticommutator)  operates. Thus, the SUSY transformations  are distinctly different from the 
(anti-) BRST symmetry transformations. The clinching point of {\it difference} is the property of  absolute anticommutativity   
[which is respected by the (anti-)BRST symmetry transformations {\it but} violated by the SUSY transformations].

In a set of research papers (see, e.g. [5,6] and references therein), we have established that any arbitrary {\it Abelian} {\it p}-form
$(p = 1,2,3...)$ 
gauge theory would respect, in addition to the (anti-)BRST symmetry transformations, the (anti-)co-BRST symmetry transformations, 
too, in the $ D = 2p$ dimensions  of  spacetime at the {\it quantum} level. This observation has been shown to be true [7] in the cases of
(1+1)-dimensional (2D) (non-)Abelian $1$-form gauge theories (without any interaction with matter fields).  In fact, these 2D theories 
have been shown [7] to be the  perfect field theoretic examples of  Hodge theory as well as a {\it new} model of topological field theory (TFT) which 
captures some salient features of Witten-type TFTs [8] as well as a few key aspects of Schwarz-type TFTs [9]. In a recent set of
couple of papers [10,11], we have discussed the Lagrangian densities, their symmetries and Curci-Ferrari (CF)-type restrictions
for the 2D {\it non-Abelian} 1-form gauge theory within the framework of BRST and superfield formalisms. 
Some novel features have been pointed out, too, in our earlier works [10, 11].

In our earlier work [10], we have been able to show the {\it equivalence} of the coupled Lagrangian densities w.r.t.
 the (anti-)BRST as well as (anti-)co-BRST symmetries of the 2D non-Abelian $1$-form gauge theory 
(without any interaction with matter fields). However, we have {\it not} been able to compute the conserved
currents (and corresponding charges) for the above continuous symmetries. One of the central themes of our present investigation
is to compute {\it all} the conserved charges and derive their algebra to show the validity of CF-type restrictions 
at the {\it algebraic} level. This exercise  establishes the {\it independent} existence of  a set of CF-type
restrictions for the 2D non-Abelian $1$-form theory (which have been shown from the symmetry considerations [10] as well as from the 
point of view of the superfield approach to  BRST formalism [11]). In our present endeavor,
we accomplish this goal in a straightforward fashion and show that the CF-type restrictions,
corresponding to the (anti-)co-BRST symmetries, have some {\it novel} features that are  completely different from the {\it usual} 
CF-condition [12] corresponding to the (anti-)BRST symmetry transformations of our present theory.

One of the highlights of our present investigation is the derivation of the CF-type restrictions and some of the equations
of motion (EOM)
from the algebra of conserved charges where the ideas of symmetry generators corresponding to the continuous symmetry
transformations of our 2D non-Abelian theory  are  exploited. Thus, to summarize the key results of our previous works [10,11] 
and {\it present} one, 
we would like to state that we have been able to show the existence of the CF-type of restrictions 
from the point of view of symmetries of the $2D$ non-Abelian 1-form gauge  theory [10], superfield approach to BRST formalism applied 
to the above $2D$ theory [11]
{\it and} algebra of the  conserved charges of the above theory. The latter (i.e. algebra) is reminiscent of the algebra 
of the de Rham cohomological operators of the differential geometry. Our present studies establish the 
{\it independent} nature of the CF-type restrictions in the context of nilpotent (anti-)co-BRST 
symmetries (existing {\it only } in the 2D non-Abelian 1-form gauge theory)  
which are {\it different} from the CF-condition [12] that appears in the context of 
(anti-)BRST symmetries (existing  in {\it any} arbitrary dimension of 
spacetime).

In our present endeavor, we have demonstrated that the {\it usual} coupled Lagrangian densities (1) (see below)
 for the non-Abelian 1-form
gauge theory respect {\it four} perfect symmetries individually whereas the 
{\it generalized} versions of these Lagrangian densities (26) (see below)
respect {\it five} perfect symmetries individually. It has been shown that {\it both} 
the Lagrangian densities of Eq. (26) respect (anti-)co-BRST
symmetries that have been listed in (27) (see below) which is a completely novel observation [cf. Eq. (28)]. 
The absolute anticommutativity of the (anti-)co-BRST charges [that have been computed from the Lagrangian densities (1)]
requires the validity of the CF-type restrictions $({\cal B}\times C = 0, {\cal B}\times \bar C = 0)$. However, 
the requirement of the absolute anticommutativity of the above charges [that are computed 
from the generalized Lagrangian densities (26)] turn out to be {\it perfect}.
This happens because of the fact that   the conditions ${\cal B}\times C = 0$ and  
${\cal B}\times \bar C = 0$ become equations of motion
for the Lagrangian densities (26). This is also a {\it novel}  observation in our present endeavor
(connected with the $2D$ non-Abelian theory).

Our present endeavor is propelled by the following key considerations.
First and foremost, we have derived the conserved charges corresponding
to the continuous symmetries which have {\it not} been discussed in our earlier works
[10,11]. Second, we have derived the CF-type  restrictions in the context of 
2D non-Abelian theory which emerge from the symmetry considerations [10] 
as well as from the application of augmented version of superfield approach to BRST
formalism [11]. We show, in our present endeavor, the existence of {\it such}
 restrictions in the language of algebra,
connected with the conserved charges, which 
obey the algebra of cohomological operators of differential geometry.
Third, the (anti-)co-BRST symmetries {\it absolutely} anticommute 
{\it without} use of any kinds of the CF-type restrictions [which is {\it not} the case with the (anti-)BRST symmetries]. 
However, in our present  endeavor, we have shown that CF-type restrictions
$({\cal B}\times C = 0, {\cal B}\times\bar C = 0)$ appear when we consider the requirement of 
absolute anticommutativity of the  (anti-)co-BRST charges [derived 
from the Lagrangian densities (1)]. Finally, we speculate that the understanding and 
insights, gained in the context of 2D non-Abelian theory, might turn out to be useful
for the 4D Abelian 2-form  and 6D Abelian 3-form gauge theories which have {\it also}
been shown to be the models of Hodge  theory [6].

The material of our present  theoretical work is  organized as follows. In Sec. 2, we {\it briefly} recapitulate the bare essentials of  
the nilpotent (anti-)BRST and (anti-)co-BRST symmetries, a {\it unique} bosonic symmetry and a ghost-scale symmetry of 
the 2D non-Abelian gauge theory in the Lagrangian formulation. Our Sec. 3 contains the details of the derivation
 of conserved Noether currents and conserved charges corresponding to the above continuous 
symmetries. Our Sec. 4 deals with the elaborate proof of the coupled Lagrangian densities to be 
{\it equivalent} \,w.r.t. the nilpotent (anti-)BRST as well as (anti-)co-BRST symmetry transformations. In Sec. 5,  we derive the
 algebraic structures of the symmetry operators and conserved charges and 
establish their connection with the cohomological operators of differential geometry (at the algebraic level). Our Sec. 6 deals with the 
discussion of some {\it novel} observations in the context of algebraic structures. Finally, we make some concluding 
remarks and point out a few future  directions  in Sec. 7.

In our Appendices
A and B, we collect some of the explicit computations that have been incorporated  into the main body of the 
text of our present endeavor. In our Appendix C, we show the consequences of the (anti-)BRST symmetry transformations
when they are applied on the generalized forms of the Lagrangian densities [cf. Eq. (26) below] where the CF-type restrictions
have been incorporated.\\

\noindent
{\it Convention and Notations}: Our whole discussion is based on the choice of the 2D
 flat metric $\eta_{\mu\nu}$ with signatures $(+1,-1)$ which corresponds to the background Minkowskian 
 2D spacetime manifold. We choose the 2D Levi-Civita tensor $\varepsilon_{\mu\nu}$ such 
 that $\varepsilon_{01} =+1=\varepsilon^{10}$ and $\varepsilon_{\mu\nu}\,\varepsilon^{\mu\nu} =-\; 2!$, $\varepsilon_{\mu\nu}\,\varepsilon^{\nu\lambda} =\delta ^{\lambda}_{\mu}$, etc. Throughout the whole body of our text,
 \,we adopt the notations for the (anti-)BRST and (anti-)co-BRST transformations
 as $s_{(a)b}$ and\,$ s_{(a)d}$, respectively. In the 2D Minkowskian flat spacetime, the field strength tensor: 
$F_{\mu\nu} = \partial_{\mu}A_{\nu} -\partial_{\nu}A_{\mu}  + i\,(A_{\mu}\times A_{\nu})$
 has only {\it one} existing component 
$E = F_{01} = -\varepsilon^{\mu\nu}[\partial_{\mu}A_{\nu} +\,\frac{i}{2}(A_{\mu}\times A_{\nu})]$
 and our Greek indices \,${\mu}\,,{\nu}\,,{\lambda}\,... = 0\,,1$ correspond to the time and space directions.
 We have also 
 adopted the dot and cross products in the\,$ SU(N)$ Lie algebraic space where
$P\cdot Q = P^a\,Q^a$ and $(P\times Q)^a = f^{abc}P^b\,Q^c$ for the non-null vectors  $P^a $
$(P = P^a T^a\equiv P\cdot T)$ and\, $Q^a$
$( Q = Q^aT^a\equiv Q\cdot T )$
where the $SU(N)$ Lie algebra is: $[T^a,T^b] = f^{abc} T^c$. In this specific mathematical algebraic relationship, 
$T^a$ are the generators of the $SU(N)$ Lie algebra
and the structure constants $f^{abc}$ are chosen to be totally antisymmetric in {\it all} their indices $a,b,c  = 1,2.....N^2-1$.\\

\noindent
{\it Standard Definition}: On a compact manifold without a boundary, the set of three mathematical operators $(d, \delta, \Delta)$
is called as a set of the de Rham cohomological operators of differential geometry where $(\delta)d$ are christened as the
(co-)exterior derivatives and $\Delta = (d + \delta)^2$ is called as the Laplacian operator. Together, these operators
satisfy an algebra: $ d^2 = \delta^2 = 0, \;\Delta = d \delta +   \delta d, \;[\Delta, d ] = 0, \; [\Delta, \delta] = 0$
which is popularly known  as the Hodge algebra of differential geometry. The 
co-exterior derivative $\delta$ and exterior derivative  $d$ are connected by
a relationship $\delta = \pm \, * \; d \;*$ where $*$ is the Hodge duality operation (defined on the given compact
manifold without a boundary). It is obvious that the (co-)exterior derivatives are nilpotent of oder two and Laplacian
operator is like the Casimir operator for the whole algebra. However, the latter (i.e. the Hodge algebra) is {\it not}
a Lie algebra.  \\

\section
{\bf Preliminaries: Lagrangian Formulation}

We begin with the coupled (but equivalent) Lagrangian densities [13,14,10,11] of our 2D non-Abelian 1-form gauge theory 
in the Curci-Ferrari gauge (see, e.g. [15,16]) as
\begin{eqnarray}
&&{\cal L}_B = {\cal B} {\cdot E} - \frac {1}{2}\,{\cal B} \cdot {\cal B} +\, B\cdot (\partial_{\mu}A^{\mu}) 
+ \frac{1}{2}(B\cdot B + \bar B \cdot \bar B) - i\,\partial_{\mu}\bar C \cdot D^{\mu}C, \nonumber\\
&&{\cal L}_{\bar B} = {\cal B} {\cdot E} - \frac {1}{2}\,{\cal B} \cdot {\cal B} - \bar B\cdot (\partial_{\mu}A^{\mu}) 
+ \frac{1}{2}(B\cdot B + \bar B \cdot \bar B) - i\, D_{\mu}\bar C \cdot \partial^{\mu}C,
\end{eqnarray}
where $B$, $\bar B $ and ${\cal B}$  are the auxiliary fields, 
 $D_{\mu} C = \partial_{\mu} C + i\, (A_{\mu}\times C) $ and $D_{\mu}\bar C = \partial_{\mu}\bar C + i\, (A_{\mu}\times\bar C)$
 are the covariant derivatives on the ghost and anti-ghost fields, respectively. These derivatives are in the {\it adjoint} 
representation of the $ SU(N)$ Lie algebra and
 ${B + \bar B + (C\times \bar C)} = 0 $  is the Curci-Ferrari (CF) condition [12]. 
The latter is responsible for the\,{\it equivalence} of the Lagrangian densities 
${\cal L}_B $ and ${\cal L}_{\bar B}$. This observation is one of the\,{\it inherent} properties of the basic concept behind the existence of {\it coupled} Lagrangian densities for a given gauge theory [13,14]. 
The fermionic $[(C^a)^2 = 0,\,  {({\bar C}^a)^2} = 0\, ]$ 
(anti-) ghost fields $(\bar C^a)C^a $ are  needed for the validity of unitarity in the theory and they  satisfy: 
$C^a \bar C^b+ \bar C^b C^a = 0, C^a C^b+C^b C^a= 0,
\bar C^a\bar C^b+\bar C^b\bar C^a = 0, \bar C^a C^b+C^b\bar C^a= 0,$ etc. We would like to remark here
that the 2D kinetic term [i.e. $- (1/4) F^{\mu\nu} \cdot F_{\mu\nu} = (1/2) E \cdot E 
\equiv {\cal B} {\cdot E} - (1/2)\,{\cal B} \cdot {\cal B}$] has been lineared by invoking the auxiliary field ${\cal B}$.

The Lagrangian densities in (1) respect the following off-shell nilpotent 
$(s_{(a)b}^2= 0)$ (anti-)BRST symmetry transformations $s_{(a)b}$:
\begin{eqnarray}
&&s_{ab} A_\mu= D_\mu\bar C,\,\,\,\,\,  s_{ab} \bar C= -\frac{i}{2}\,(\bar C\times\bar C),
\,\,\,\, s_{ab}C   = i{\bar B},\,\,\,\,  s_{ab}\bar B    = 0
,\,\,\,\, s_{ab}({\cal B}\cdot{\cal B}) = 0,\nonumber\\
&&s_{ab} E      = i \,(E\times\bar C),\,\;\;
\,\,\; s_{ab}{\cal B} =  i\,({\cal B}\times\bar C),
\,\,\, \; \;s_{ab} B = i\,(B \times \bar C),\quad \;s_{ab}({\cal B}\cdot E) = 0, \nonumber\\
&&s_b A_\mu = D_\mu C,  \;\,\,\,\,s_b C =  - \frac{i}{2} (C\times C),\,\,\,\;  s_b\bar C \;= i\,B ,\;
 \,\,\,\, \;s_b B = 0,\,\,\,\, \;s_b({\cal B}\cdot{\cal B}) = 0, \nonumber\\
&& s_b\bar B = i\,(\bar B\times C),\;\;\,\,\,s_b E = i\,(E\times C),\qquad \,\,\,s_b {\cal B} 
= i\,{(\cal B}\times C),\quad \,\,\,
s_b({\cal B}\cdot E) = 0.
\end{eqnarray}
This is due to the fact that we observe the following:
\begin{eqnarray}
&&s_b{\cal L}_B = \partial_\mu(B \cdot D^\mu C),  
\qquad\qquad\qquad\quad s_{ab}{\cal L}_{\bar B}= - \;\partial_\mu{(\bar B \cdot D^\mu \bar C)}.
\end{eqnarray}
As a consequence, the (anti-)BRST transformations are the {\it symmetry} transformations for the action integrals 
${S= \int d^2x \,{\cal L}_B}$ and ${S = \int d^2x \, {\cal L}_{\bar B}}$, respectively. 
The (anti-)BRST symmetry transformations absolutely anticommute with each other (i.e. 
${\{s_b,s_{ab}}\} = 0$) {\it only}  when the CF-condition is satisfied. One of the decisive features of the
(anti-)BRST symmetry transformations is the observation that 
the kinetic term ($ -\frac{1}{4}F_{\mu\nu}\cdot F^{\mu\nu}$ =$\frac{1}{2}E\cdot E\equiv {\cal B}\cdot E$ - $\frac{1}{2}
{{\cal B}}\cdot{{\cal B}}$) remains invariant under it.
 This observation would be exploited, later on, in establishing 
a connection between the continuous symmetries of our 2D  theory 
and cohomological operators of differential geometry at the {\it algebraic} level.

In addition to the (anti-)BRST symmetry transformations (2), we note the presence of the following nilpotent 
$(s_{(a)d}^2 = 0) $ and absolutely anticommuting $(s_d s_{ad}+s_{ad} s_d = 0)$ (anti-)co-BRST symmetry 
transformations in the theory (see, e.g. [7] for details): 
\begin{eqnarray}
&&s_{ad} A_\mu = - \varepsilon_{\mu\nu}\partial^\nu C,\quad\, s_{ad} C = 0,\qquad\quad\,\,\,\,\,\,
s_{ad} \bar C =  i {\cal B},\qquad\qquad\,s_{ad} {\cal B} = 0,\nonumber\\
&& s_{ad} E =D_\mu\partial^\mu C,\quad\quad\,\,\, s_{ad} B = 0,\qquad\qquad
s_{ad}\bar B = 0,\qquad\, s_{ad}({\partial_\mu A^\mu})= 0, \nonumber\\
&&s_d A_\mu = - \varepsilon_{\mu\nu}\partial^\nu \bar C,
\quad\quad s_d C = - i {\cal B},\quad\qquad s_d \bar C = 0,\,\qquad\qquad\quad s_d{\cal B} = 0,\nonumber\\
&& s_d E = D_\mu\partial^\mu\bar C,\,\qquad\quad s_d B = 0,
\qquad\qquad\,\, s_d\bar B = 0, \qquad\quad s_d({\partial_\mu A^\mu})= 0. 
\end{eqnarray}
The Lagrangian $ {\cal L}_B $ and  ${\cal L}_{\bar B}$ transform, under the above transformations, as follows
\begin{eqnarray}
&&s_{ad}\,{\cal L}_{\bar B} = \partial_\mu[{\cal B}\cdot \partial^\mu C],  \qquad\qquad\qquad\quad s_d {\cal L}_B = \partial_\mu[{\cal B}
\cdot\partial^{\mu} \bar C],
\end{eqnarray}
which imply that the action integrals ${S = \int d^2x \,{\cal L}_{B}}$ and ${ S = \int d^2x \, {\cal L}_{\bar B }}$ remain 
invariant under the (anti-)co-BRST  transformations. 
One of the decisive features of the (anti-)co-BRST symmetries is the
observation that the gauge-fixing term $( \partial_\mu A^\mu) $  remains invariant under them. This observation would play a
key role in establishing a connection between these symmetries and the cohomological operators of differential geometry at the {\it algebraic} level. It is quite clear that
we have {\it four}  fermionic symmetries in our present 2D  theory.

There are {\it two} bosonic symmetries in our theory, too. The {\it first}  one is the ghost-scale symmetry $(s_g)$ and  {\it second} 
one is a unique bosonic symmetry $s_w = {\{s_d,s_b}\} = -{\{s_{ad},s_{ab}}\}$. We focus $first$ on the ghost-scale symmetry.
Under this symmetry, we have the following transformations  for the fields of our present theory, namely;
\begin{eqnarray}
&& C\longrightarrow e^\Omega \,C,\qquad {\bar C}\longrightarrow  e^{-\Omega }\,{\bar C},\qquad \Phi\longrightarrow e^0\,{\Phi},
\end{eqnarray}
where the generic field $\Phi = A_{\mu},\,B,\,{\cal B},\,\bar B,\,E $ and $\Omega$ is a {\it global} \,(spacetime independent)\,scale transformation parameter. One of the decisive features of the ghost-scale symmetry transformations
is the observation that {\it only} the (anti-)ghost
fields transform and the remaining ordinary basic/auxiliary fields of the theory remain {\it invariant} under them.
The infinitesimal version $(s_g)$ of the above ghost-scale symmetry transformations is:
\begin{eqnarray}
&& s_g C = C,\qquad s_g{\bar C}  = -{\bar C},\qquad s_g{\Phi }= 0.
\end{eqnarray}
In  the above, we have set $\Omega  = 1$ for the sake of brevity. Under these infinitesimal transformations, it can be 
readily checked that: 
\begin{eqnarray}
&&s_g {\cal L}_B = 0,\qquad\qquad\qquad s_g{\cal L}_{\bar B} = 0.
\end{eqnarray}
Thus, the action integrals automatically remain invariant under the above ghost-scale symmetry transformations. Now, we focus on
the bosonic symmetry ${s_w }$ of our theory. It is elementary to check that, for the Lagrangian density ${\cal L}_B $, we
have
\begin{eqnarray}
&&s_w A_\mu = -[ D_\mu{\cal B} + \varepsilon_{\mu\nu} \,(\partial^\nu\bar C\times C) + \varepsilon_{\mu\nu}\,\partial^\nu B],
\qquad s_w \bar B = (\bar B\times {\cal B}),\nonumber\\
&&s_w(\partial_{\mu} A^{\mu}) = -[\partial_{\mu} D^{\mu}{\cal B} + \varepsilon_{\mu\nu}(\partial^{\nu}\bar C\times \partial_{\mu} C)],
\qquad s_w [C,\bar C,{\cal B},B] = 0,   \nonumber\\
&&s_w E = -[D_{\mu}\partial^{\mu}{\cal B} + i\,(E\times{\cal B}) - D_{\mu } C\times\partial^{\mu}\bar C- D_{\mu}\partial^{\mu}\bar C\times C],
\end{eqnarray}
where we have taken $ s_w = {\{s_b,s_d}\}$ (modulo a factor of {\it i}) and $E = -\varepsilon^{\mu\nu}(\partial_{\mu}A_{\nu} +\frac{i}{2}A_{\mu}\times A_{\nu})$. One of the key observations is  that the (anti-)ghost fields of the theory {\it do not} transform under the bosonic
symmetry transformation $s_w$.
It can be checked that the Lagrangian density ${\cal L}_B$ 
transforms under this bosonic symmetry transformation\footnote{There is a simple  way  to derive Eq. (10). Using the basic definition $s_w = {\{s_b,s_d}\} $ and applying it on ${\cal L}_B$ we obtain Eq. (10) [with the inputs from (3) and (5)].}
 as (see, e.g. [10])
\begin{eqnarray}
&&s_w{\cal L} _B = \partial_{\mu}[{\cal B}\cdot\partial^{\mu}B-B\cdot D^{\mu}{\cal B}-\partial^{\mu}\bar C\cdot({\cal B}\times C)
- \varepsilon^{\mu\nu} B\cdot(\partial_{\nu}\bar C\times C)],
\end{eqnarray}
thereby rendering the action integral $ S =\int d^2 x\,{\cal L}_B$ invariant. 
Thus, the bosonic transformations (9) correspond to the {\it symmetry} of the theory
\footnote{The unique bosonic symmetry transformation $s_w = {\{s_b, s_d\}}$
 is independent of $s_b$ as well as $s_d$
because it commutes (i.e. $[s_w, s_b] = [s_w, s_d] = 0$) with both of them [cf. Eq. (29) below]}. 
We remark that one can define another bosonic symmetry $s_{\bar w} = -\;{\{s_{ad},\,s_{ab}}\}$
for the Lagrangian density ${\cal L}_{\bar B} $ but it turns out to 
be equivalent (i.e.\;$ s_w + s_ {\bar w }=0$) to $s_w = {\{s_d,\,s_b}\}$ 
if we use the equations of motion of the theory and the CF-type condition
\,$(B+\bar B+( C \times \bar C ) = 0)$. To sum up, we have total\,{\it six }
continuous symmetries in the theory. Together,\,these symmetry operators satisfy
an algebra that is\,{\it exactly} similar to the algebraic structure 
of the cohomological operators of the differential geometry. 
Thus, there is a connection between the {\it two} (cf. Sec. 5 below).

\noindent
\section
{\bf Conserved Charges: Noether Theorem}

\noindent
The Noether theorem states that the invariance of the action integral, under continuous symmetry transformations,
leads to the existence of conserved currents. As pointed out earlier, the Lagrangian densities
${\cal L}_B$ and $ {\cal L}_{\bar B}$ transform, under $s_b$ and $s_{ab}$, to the total spacetime derivatives as given in (3)
thereby rendering the action integrals $S =\int d^2x\,{\cal L}_B$ and $S =\int d^2x\,{\cal L}_{\bar B}$ invariant. 
The corresponding Noether currents (w.r.t. BRST and anti-BRST symmetry transformations) are:
\begin{eqnarray}
&& J^{\mu}_b = -\varepsilon^{\mu\nu}{\cal B}\cdot D_{\nu} C 
+ B\cdot D^{\mu} C+\frac{1}{2}\,
\partial^{\mu}\bar C\cdot(C\times C),\nonumber\\
&&J^{\mu}_{ab} = -\varepsilon^{\mu\nu}{\cal B}\cdot D_{\nu}\bar C 
-\bar B\cdot D^{\mu}\bar C
-\frac{1}{2}\,\partial^{\mu} C\cdot(\bar C\times\bar C).
\end{eqnarray}
The above currents are conserved (i.e. $\partial_{\mu}J^{\mu}_b = 0$ and $\partial_{\mu} J^{\mu}_{ab} = 0$) due
 to the following Euler-Lagrange (EL) equations of motion\,(EQM) that emerge from 
${\cal L}_B$ and ${\cal L}_{\bar B}$, namely;
\begin{eqnarray}
&& {\cal B} = E,\quad D_\mu\partial^{\mu}\bar C = 0,\qquad \partial_{\mu} D^{\mu} C = 0,\qquad 
\varepsilon^{\mu\nu}D_{\nu}{\cal B} + \partial ^{\mu}B+(\partial^{\mu}\bar C\times C) = 0,\nonumber\\
&& {\cal B} = E,\quad \partial_\mu D^{\mu}\bar C = 0,\qquad D_{\mu}\partial^{\mu}C = 0,\; \varepsilon^{\mu\nu}D_{\nu}{\cal B} -
 \partial ^{\mu}\bar B-({\bar C}\times \partial^{\mu} C) = 0.
\end{eqnarray}
The above observations are sacrosanct as far as Noether's theorem is concerned. It is to be noted that we have used $E = -\varepsilon^{\mu\nu}(\partial_{\mu}A_{\nu}+\frac{i}{2} A_{\mu}\times A_{\nu})$ in the derivation of the EOM.

The conserved charges (that emerge out from the   Noether currents) are:
\begin{eqnarray}
&&Q_b =\int dx \;J^0_b\equiv \int dx \;[{\cal B}\cdot D_1 C + B\cdot D_0 C 
+\frac{1}{2}\dot{\bar C}\cdot(C\times C)],\nonumber\\
&&Q_{ab} =\int dx\; J^0_{ab}\equiv \int dx \;[{\cal B}\cdot D_1\bar C 
- \bar B\cdot D_0\bar C -\frac{1}{2}\cdot(\bar C\times\bar C)\cdot\dot C].
\end{eqnarray}
Using the EL-EOM (12), the above charges can be expressed in a more useful (but equivalent) forms as 
\begin{eqnarray}
&&Q_b =\int dx \;[B\cdot D_0 C -\dot B\cdot C -\frac{1}{2}\,\dot{\bar C}\cdot(C\times C)],\nonumber\\
&&Q_{ab} =\int dx\;[\dot{\bar B}\cdot\bar C -\bar B\cdot D_0\bar C +\frac{1}{2}(\bar C\times \bar C)\cdot\dot C],
\end{eqnarray}
which are the {\it generators} for the (anti-)BRST transformations\;(2). This statement can be verified by observing that 
the (anti-)BRST symmetry transformations, listed in equation (2), can be derived  from 
the following general expression
\begin{eqnarray}
s_r \, {\Phi} = \mp \, i\,\,[\Phi, Q_r]_{\mp}\qquad\qquad\qquad r = b, ab,
\end{eqnarray}
where the subscripts ($\mp$), on the square bracket, correspond to the bracket being 
commutator and anticommutator for the generic field $\Phi$ being bosonic and fermionic, respectively. 
The signs $\mp$ in front of square bracket can be chosen appropriately (see, e.g. [17] for details).

Under the (anti-)co-BRST transformations $s_{(a)d}$,\, the\, Lagrangian\, densities \,${\cal L}_B$ 
and ${\cal L}_{\bar B}$   transform as given in (5).
According to the Noether theorem, these infinitesimal continuous  transformations 
lead to the derivation of conserved Noether currents. The explicit 
expressions for these conserved currents are:
\begin{eqnarray}
&&J^{\mu}_d = {\cal B}\cdot\partial^{\mu}\bar C -\varepsilon^{\mu\nu} B\cdot\partial_{\nu}\bar C,\qquad\qquad
J^{\mu}_{ad} = {\cal B}\cdot\partial^{\mu} C +\varepsilon^{\mu\nu}\bar B\cdot\partial_{\nu} C.
\end{eqnarray} 
The conservation laws $\partial_{\mu}J^{\mu}_d = 0$ and $\partial_{\mu}J^{\mu}_{ad} = 0$ can be proven by 
using EL-EQM (12). The conserved charges can be expressed\,{\it equivalently }in various forms as: 
\begin{eqnarray}
&&Q_d =\int\, dx \;J^0_d =\int dx\;[{\cal B}\cdot\dot{\bar C}+B\cdot\partial_1\bar C]
\equiv \int dx\;[{\cal B}\cdot\dot{\bar C}-\partial_1 B\cdot\bar C]\nonumber\\
&&\qquad\qquad\qquad\;\;\equiv  \int dx\;[{\cal B}\cdot\dot{\bar C}-D_0{\cal B}\cdot\bar C 
+(\partial_1\bar C\times C)\cdot\bar C],\nonumber\\
&&Q_{ad} =\int dx \;J^0_{ad} =\int dx\;[{\cal B}\cdot\dot C- \bar B \cdot\partial_1 C]
\equiv \int dx\;[{\cal B}\cdot C +\partial_1 \bar B\cdot C]\nonumber\\
&&\qquad\qquad\qquad\quad\, \equiv \int dx\;[{\cal B}\cdot\dot C-D_0{\cal B}\cdot C -(\bar C\times \partial_1 C)\cdot C].
\end{eqnarray}
The above charges are the generators of the (anti-)co-BRST symmetry transformations in equation (4). This statement can be
corroborated by using the formula (15) where we have to replace:\,$r= a,ab\longrightarrow $ $r = d,ad $.

We remark that the  fermionic symmetries $s_{(a)b}$ and $s_{(a)d}$ are 
off-shell nilpotent of order two (i.e. $ \,s_{(a)b}^2 = 0$,\,\,$s_{(a)d}^2 = 0$). This can be explicitly checked  from the 
transformations listed in equations\,(2) and (4). This property (i.e. nilpotency) is also reflected at the level of 
conserved charges. To corroborate this assertion, we note that
\begin{eqnarray}
&&s_b Q_b = -i\,{\{Q_b,Q_b}\} = 0\qquad\qquad \Longrightarrow \quad Q_b^2 = 0,\nonumber\\
&&s_{ab} \,Q_{ab} = -i\,\{Q_{ab},Q_{ab}\} = 0 \quad\quad\,\Longrightarrow \quad Q_{ab}^2 = 0,\nonumber\\
&&s_d Q_d  = -i\,{\{Q_d,Q_d}\} = 0 \qquad\,\,\,\,\quad\Longrightarrow \quad Q_d^2 = 0,\nonumber\\
&&s_{ad}Q_{ad} = -i\,{\{Q_{ad},Q_{ad}}\} = 0\quad\quad \,\Longrightarrow \quad Q_{ad}^2 = 0,
\end{eqnarray}
where we have used the definition of the symmetry generator (15). This observation is straightforward because the l.h.s.
of the above equations can be computed explicitly by using the expressions for $ Q_{(a)b}$,  $Q_{(a)d}$ [cf. Eqs.\,(14) 
and (17)] and the transformations (2) and (4) corresponding
to the (anti-)BRST and (anti-)co-BRST  continuous symmetry transformations\footnote{ These claims are true for any arbitrary
expressions for the charges listed in (13), (14) and (17)
 provided we take into account the symmetry transformations (2) and (4).}.

The conserved Noether current and corresponding charge for the infinitesimal and continuous ghost-scale transformations\,(7) are:
\begin{eqnarray}
&&J^{\mu}_g = - i\,[\partial^{\mu}\bar C\cdot C -\bar C\cdot D^{\mu } C],\nonumber\\
&&Q_g =\int dx \;J^0_g\equiv  - i\int dx \;[\dot{\bar C}\cdot C -{\bar C}\cdot D_0\ C].
\end{eqnarray}
Using the equations of motion (12), it can be readily checked $\partial_{\mu}J^{\mu}_g=0$. Hence, the charge $Q_g $
is also conserved. Finally, we discuss a bit about the {\it unique} bosonic symmetry 
transformations $ s_w = {\{s_d,s_b}\} = -{\{s_{ad},s_{ab}}\}$ in this theory [7]. As pointed out earlier,
the Lagrangian density ${\cal L}_B $ transforms to the total spacetime derivative 
under $s_w$ as given in (10). The conservation of  Noether current  
(i.e. $\partial_{\mu}J^{\mu}_w = 0) $ can be proven by using Eq.\,(12). The conserved  
current $(J^{\mu}_w)$ and corresponding charge\,$(Q_w)$ are [7]:
\begin{eqnarray}
&&J^{\mu}_w =-\varepsilon^{\mu\nu}[{\cal B}\cdot D_{\nu}{\cal B} -B\cdot\partial_{\nu}B],\nonumber\\
&&Q_w =\int dx \;J^0_w =\int dx\; [{\cal B}\cdot D_1{\cal B} -B\cdot\partial_1 B].
\end{eqnarray}
It our Appendix B, we have shown the alternative derivations of $ Q_w $ from the continuous symmetry transformations
and the  concept behind the symmetry generator.
It is evident that we have  {\it six} conserved charges  which correspond to the {\it six}
infinitesimal  and continuous 
symmetries that exist in our theory. We shall establish their connections with the de Rham 
cohomological operations of differential geometry in our Sec. 5 where the emphasis would be laid  on the
algebraic structure(s) {\it only}.

\noindent
\section{\bf Equivalence of the Coupled Lagrangian  Densities:  Continuous Symmetry Considerations}

We observe, first of all, that ${\cal L}_B$ and ${\cal L}_{\bar B}$ are equivalent  
{\it only }when the CF-condition  ${B + \bar B + (C\times \bar C)} = 0 $ is satisfied. 
This can be shown by the requirement 
of the equivalence of the Lagrangian densities (i.e  ${\cal L}_B$ - ${\cal L}_{\bar B} \equiv  0$, modulo a total spacetime derivative term) 
which primarily leads to the following equality, namely; 
\begin{eqnarray}
&&B\cdot(\partial_{\mu} A^{\mu}) - i\,\partial_{\mu}\bar C\cdot D^{\mu} C = -\bar B\cdot(\partial_{\mu} A^{\mu}) - i\,
D_{\mu}\bar C\cdot \partial^{\mu} C.
\end{eqnarray}
 Thus, it is evident that {\it both} 
the Lagrangian  densities are {\it equivalent} only on a space of quantum fields  which is 
defined by the CF-condition (i.e. ${B + \bar B + (C\times \bar C)} = 0 )$ in the 2D Minkowskian flat spacetime manifold.  Furthermore,
we note that {\it both} the  Lagrangian  densities respect the (anti-)BRST symmetry  
transformations  because, we observe that, besides  (3), we have the following explicit transformations:  
\begin{eqnarray}
&&s_{ab}{\cal L}_B = -\partial_\mu\,[{\{\bar B + ( C\times\bar C)\} \cdot \partial^\mu \bar C}\,] 
+\{(B+\bar B + ( C \times {\bar C})\} \cdot D_\mu \partial^\mu \bar C, \nonumber\\
&&s_b{\cal L }_{\bar B}\; = \partial_\mu\,[ {\{ B + ( C \times \bar C )\}}\cdot
\partial^\mu C \,]-{\{B + \bar B + ( C\times\bar C )\}}\cdot D_\mu\partial^\mu C.
\end{eqnarray}
Thus, if we exploit the strength of the CF-condition: ${B + \bar B + (C\times \bar C)} = 0, $  we obtain
the following symmetry transformations, namely;
\begin{eqnarray}
&&s_{ab}\,{\cal L}_{B} = \partial_\mu[B \cdot\partial^{\mu}\bar C ],  \qquad\qquad 
s_b{\cal L}_{\bar B} = -\partial_\mu[{\bar B}\cdot\partial^{\mu}C],
\end{eqnarray}
thereby rendering the action integrals invariant. We draw the conclusion that, 
due to the key equations (3) and (23), {\it both} the Lagrangian  densities ${\cal L}_B$ and 
${\cal L}_{\bar B}$ respect {\it both} the BRST and anti-BRST symmetries 
provided we confine ourselves on the space of quantum fields in the Hilbert space defined by the CF-condition (where the absolute
anticommutativity property (i.e. ${\{s_b, s_{ab}}\} = 0 $) is {\it also} satisfied for $s_{(a)b}$). 
As a consequence, we infer that {\it both} 
the Lagrangian  densities are {\it equivalent} w.r.t. the (anti-)BRST symmetries
on the space of quantum fields in the Hilbert space defined by the CF-condition [12].
Now we focus on the issue of\,{\it equivalence} of the Lagrangian densities ${\cal L}_B$ and ${\cal L}_{\bar B}$ from
the point of view of the\,(anti-)co-BRST symmetry transformations. Besides the symmetry transformation in
equation (5), we observe the following: 
\begin{eqnarray}
&& s_d{\cal L}_{\bar B} = \partial_\mu[{\cal B}\cdot D^\mu\bar C- \varepsilon^{\mu\nu} (\partial_\nu\bar C\times  
\bar C)\cdot C] + i\; (\partial_\mu A^\mu)\cdot({\cal B}\times\bar C),\nonumber\\
&& s_{ad}{\cal L}_B = \partial_\mu[{\cal B}\cdot D^\mu C+ \varepsilon^{\mu\nu}\bar C\cdot(\partial_\nu C\times C)] 
+ i\; (\partial_\mu A^\mu)\cdot({\cal B}\times C).
\end{eqnarray}
We draw the conclusion, from the above, that {\it both} the Lagrangian  densities 
${\cal L}_B$ and ${\cal L}_{\bar B}$ are {\it equivalent}  w.r.t. the (anti-)co-BRST symmetry 
transformations if and only if the conditions 
$({\cal B}\times C) =0 $, $({\cal B}\times\bar C)=0 $ 
are satisfied. Taking the analogy with equations (22) and (23), it is  straightforward 
to conclude that ${\cal B}\times C = 0$ and  ${\cal B}\times\bar C = 0$ are the CF-type restrictions\footnote 
{We lay emphasis on the fact that these restrictions {\it do} not imply that $C\times \bar C = 0$ (thereby rendering the
theory to become Abelian). This is due to the fact that the absolute anticommutativity property
${\{s_d,s_{ad}}\}= 0$ implies that the  CF-type restrictions ${\cal B}\times C = 0$ and ${\cal B}\times \bar C = 0$
are {\it independent} of each other (see, e.g. [10] for details). In other words, {\it both} these restrictions
should be considered separately and idependently.} 
w.r.t.\, the (anti-)co-BRST symmetries for the {\it self-interacting}
 2D non-Abelian  gauge theory.

We would like to mention here that there are differences between the CF-condition
${B + \bar B + (C\times \bar C)} = 0 $ (existing for the non-Abelian 1-form gauge theory in the context of (anti-)BRST 
symmetry transformations for {\it any} arbitrary dimension of spacetime) and the CF-type restrictions that appear in the 
context of (anti-) co-BRST symmetry transformations for the 2D non-Abelian 1-form gauge theory. Whereas the latter conditions 
${\cal B}\times C = 0$ and  ${\cal B}\times\bar C = 0$ are {\it perfectly} (anti-)co-BRST invariant 
[i.e. $s_{(a)d}({\cal B}\times C) = 0$, $s_{(a)d}({\cal B}\times\bar C) = 0$] quantities, the same 
is {\it not} true in the case of CF-condition ${B + \bar B + (C\times \bar C)} = 0. $  It can be checked that: 
\begin{eqnarray}
&&s_b[{B + \bar B + (C\times \bar C)}] = i\,[{B + \bar B + (C\times \bar C)}]\times C, \nonumber\\
&&s_{ab}[{B + \bar B + (C\times \bar C)}] = i\,[{B + \bar B + (C\times \bar C)}]\times\bar C.  
\end{eqnarray}
The above transformations show that the CF-condition
$B + \bar B + (C\times \bar C)= 0$ is the (anti-) BRST
invariant quantity {\it only} on the space of quantum fields defined by the 
restriction $B + \bar B + (C\times \bar C)= 0$.
Furthermore, the (anti-)BRST symmetry transformations are absolutely anticommuting (i.e. ${\{s_b,s_{ab}}\} = 0$) {\it only} on
the  space of quantum fields defined  by the CF-condition ${B + \bar B + (C\times \bar C)} = 0 $.\,However, the absolute 
anticommutativity of the nilpotent\,(anti-)co-BRST symmetry transformations (i.e.\,${\{s_d, s_{ad}\} = 0}$) is satisfied 
$without$ any use of ${\cal B}\times C = 0$ and ${\cal B}\times\bar C = 0$. In other words, the absolute
anticommutativity of the (anti-)co-BRST symmetry transformations does not need any kinds of restrictions
from outside. We shall see, later on, that the above 
CF-type restrictions (i.e. ${\cal B}\times C = 0$ and ${\cal B}\times\bar C = 0$)
appear at the level of algebra obeyed by the conserved charges 
[derived from the Lagrangian densities (1)]
 when we demand the absolute
anticommutativity of the co-BRST and anti-co-BRST charges.

As pointed out earlier, we have seen that $s_{(a)d}({\cal B}\times C )= 0$ and 
$s_{(a)d}({\cal B}\times \bar C) = 0$. Thus, these CF-type constraints are\, (anti-)co-BRST 
invariant and, therefore, they are physical and theoretically very useful. As a consequence
of the above observation, the Lagrangian densities ${\cal L}_B $ and 
${\cal L}_{\bar B}$ can be {\it modified} in such a
manner that ${\cal L}_B $ and ${\cal L}_{\bar B}$
can have the {\it perfect} (anti-)co-BRST symmetry invariance(s). For instance, we note that the 
following modified versions of the Lagrangian densities, with fermionic 
($\lambda ^2 = \bar\lambda^2=0,\,\bar\lambda\lambda+\lambda\bar\lambda = 0)$ Lagrange
multiplier fields ${\lambda }$ and $ {\bar\lambda }$,  namely;
\begin{eqnarray}
{\cal L}^{(\lambda)}_{\bar B} &=& {\cal B} {\cdot E}-\frac {1}{2}\,{\cal B} \cdot {\cal B} - \bar B\cdot (\partial_{\mu}A^{\mu}) 
+ \frac{1}{2}(B\cdot B + \bar B \cdot \bar B)\nonumber\\
&-& i\, D_{\mu}\bar C \cdot \partial^{\mu}C + 
\lambda\cdot({\cal B}\times\bar C), \nonumber \\
{\cal L}^{(\bar\lambda)}_B &=& {\cal B}{\cdot E}-\frac {1}{2}\,{\cal B} \cdot {\cal B} 
+ B\cdot (\partial_{\mu}A^{\mu}) + \frac{1}{2}(B\cdot B 
+ \bar B \cdot\bar B)\nonumber\\ 
&-& i\,\partial_{\mu}\bar C \cdot D^{\mu}C + \bar\lambda\cdot({\cal B}\times C),
\end{eqnarray}
respect the following {\it perfect} (anti-)co-BRST symmetry transformations:
\begin{eqnarray}
&&s_{ad} A_{\mu} = - \varepsilon_{\mu\nu}\partial^\nu C,\quad\quad s_{ad} C = 0,\quad
\quad s_{ad}\,  \bar C =  i\; {\cal B},\qquad\quad s_{ad} {\cal B} = 0,\nonumber\\
&&s_{ad} E =D_\mu\partial^\mu C,\quad\quad s_{ad}({\partial_\mu A^\mu})= 0, \quad 
s_{ad}\,{\lambda}= -i\;({\partial_\mu A^\mu}),\quad s_{ad}\,{\bar \lambda}= 0, \nonumber\\
&&s_d A_\mu = - \varepsilon_{\mu\nu}\partial^\nu \bar C,
\quad\quad s_d \bar C = 0,\quad\quad\quad s_d C = - i\; {\cal B},\qquad\qquad s_d{\cal B} = 0,\nonumber\\
&& s_d E = D_\mu\partial^\mu\bar C,\qquad s_d({\partial_\mu A^\mu})= 0,\quad\quad
s_d \,{\bar \lambda} = -i\;({\partial_\mu A^\mu}),\quad\quad s_d\,{\lambda } = 0.
\end{eqnarray}
We remark here that the above (anti-)co-BRST symmety transformations are off-shell nilpotent as well as absolutely
anticommuting (without any use of CF-type restrictions). Hence, these symmetries are proper and perfect.
In the above, the superscripts $(\lambda)$ and ${(\bar\lambda)}$ 
on the Lagrangian densities are due to obvious reasons (i.e. they characterize ${\cal L}_B$ and ${\cal L}_{\bar B}$).  It should be noted that the Lagrange multipliers ${\lambda}$ and 
$\bar{\lambda}$ carry the ghost numbers equal to (+1) and (-1), respectively. 
Ultimately, we observe that the following transformations of the coupled (but equivalent) Lagrangian densities are
true, namely;
\begin{eqnarray}
&&s_d {\cal L}_B^{(\bar\lambda)} = \partial_{\mu}[{\cal B}\cdot\partial^{\mu}\bar C],
\qquad\qquad s_{ad} {\cal L}^{(\lambda)}_{\bar B} = 
\partial_{\mu}[{\cal B}\cdot\partial^{\mu} C],\nonumber\\
&&s_d {\cal L}^{(\lambda)}_{\bar B} = \partial_{\mu}[{\cal B}\cdot D^{\mu}\bar C
-\varepsilon^{\mu\nu}(\partial_\nu\bar C\times\bar C)\cdot C ],\nonumber\\
&& s_{ad}{\cal L}_B^{(\bar\lambda)} =\partial_{\mu}[{\cal B}\cdot D^{\mu}  C
+ \varepsilon^{\mu\nu} \bar C\cdot(\partial_{\nu} C\times C)].
\end{eqnarray}
which show  that the action integrals $S = \int d^2 x \, {\cal L}^{(\bar\lambda)}_B$
and $S =\int d^2 x \, {\cal L}^{(\lambda)}_{\bar B}$ remain invariant under
the (anti)co-BRST symmetry transformations $s_{(a)d}$.
Thus, we lay emphasis on the observation that $both$ the Lagrangian densities ${\cal L}_B^{(\bar\lambda)}$
and ${\cal L}_{\bar B}^{(\lambda)}$ (cf. Eq. (26)) are {\it equivalent} as far as
the $symmetry$ considerations w.r.t. the (anti-) co-BRST symmetry transformations (27) are concerned.
Henceforth, we shall $only$ focus on the Lagrangian densities ${\cal L}_B^{(\bar\lambda)}$
and ${\cal L}_{\bar B}^{(\lambda)}$ for our further discussions and we shall discuss their symmetry properties 
under the off-shell nilpotent (anti-)BRST transformations, too (cf. Appendix C below) .

\section {\bf Algebraic Structures: Symmetries and Charges}

\noindent
The Lagrangian densities in Eq. (26) are good enough to provide
the physical realizations of the cohomological operators of differential geometry
in the language of their symmetry properties. First of all, let us focus on 
${\cal L}_B^{(\bar\lambda)}$. This Lagrangian density (and corresponding
action integral) respect the (anti-)co-BRST symmetry transformations (27)
 and  BRST symmetry transformations in a
{\it perfect}  manner  because the nilpotent BRST  symmetry 
transformations $(s_b)$, listed in (2) (along with 
$s_b \,{\bar\lambda} = 0$), are the $symmetry$ of the action integral
$S =\int d^2x \,{\cal L}_B^{(\bar\lambda)}$.  This is because of the fact that we have 
$s_b({\cal B}\times C) = 0$  due to the nilpotency condition:  
$s_b^2{\cal B} = i\,s_b({\cal B}\times C) = 0$ and $s_b{\cal L}_B = \partial_{\mu}(B\cdot D^{\mu} C)$
[cf. Eq. (3)]. To be more precise, the Lagrangian density ${\cal L}_B^{(\bar\lambda)}$ respects $s_b, s_d,s_{ad},s_g$
and $s_w = {\{s_b,s_d}\}$ as discussed in Sec.  2 [with the additional transformations
$s_b{\bar\lambda} = 0$, $s_b({\cal B}\times C) = 0$ and the transformations (27) which lead to (28)].
This observation should be contrasted with the Lagrangian density ${\cal L}_B$ (cf. Eq. (11)) which respects 
only {\it four} {\it perfect} symmetries, namely; $s_b, s_d, s_g$ and $s_w$. It does not
respect $s_{ad}$ {\it perfectly}. 
One can  explicitly check that, in their
operator form, the  above set of five {\it perfect}
symmetries\footnote { We mean by the {\it perfect} symmetries as the transformations 
for which the Lagrangian densities {\it either} remain invariant {\it or} transform to the total space time derivative {\it without } any use of CF-type restrictions and/or the Euler-Lagrange EOMs.} obey the following algebra
\begin{eqnarray}
&& s_{(a)d}^2 =  s_b^2 = 0,\qquad\qquad {\{s_b,s_d}\} = s_w,\qquad {\{s_d,s_{ad}}\} = 0, \nonumber\\
&&[s_w,s_r ] = 0\qquad\quad\,\; r = b,d,ad,g,\quad{\{s_b,s_{ad}}\} = 0,\nonumber\\
&&[s_g, s_b] = + s_b,\qquad [s_g, s_d] = -s_d,\qquad\,\, [s_g,s_{ad}] = + s_{ad}.
\end{eqnarray}
In the above, we note that $s_w\bar\lambda = 0$ ($\Longrightarrow $ $ s_b\bar\lambda= 0$, $s_d\bar\lambda = 0)$
and $s_g\bar\lambda = -\bar\lambda.$
The  algebra in (29) is reminiscent of the algebra obeyed by the de Rham cohomological
operators of differential geometry (see, e.g. [18,19]), namely;
\begin{eqnarray}
&& d^2 = 0,\qquad \delta^ 2 = 0, \qquad{\{d,\delta }\} =\triangle ,\qquad[\triangle ,d] = 0=[\triangle,\delta].
\end{eqnarray}
where $(d,\delta,\triangle)$ are the exterior derivative, co-exterior derivative 
and Laplacian operators, respectively.
These operators constitute the set of de Rham cohomological operators.
It is clear that we have $ d\longleftrightarrow   s_b$\;, $ \delta \longleftrightarrow s_d$\;
and $ \triangle\longleftrightarrow s_w$. Such identification is justified due to the algebra of
the conserved charges, too, where the transformation $s_g$ and  corresponding charge $Q_g$
play an important role. We shall discuss it later. We note here that there is {\it one-to-one}
mapping between the symmetry operators and cohomological operators.

It is worth pointing out that the algebra in (29) is obeyed for the 
Lagrangian density ${\cal L}_B^{(\bar\lambda)}$ (which respects $five$ perfect continuous 
symmetries). However, the algebra (29) is satisfied $only$ 
on the on-shell where we use the  EQM (derived from Lagrangian density 
${\cal L}_B^{(\bar\lambda)}$) and the set of CF-type restrictions 
that have been discussed in earlier works [10,11]. We list here a few of these 
algebraic relationships which are juxtaposed along with the  EL-EQM and the constraints 
(i.e. CF-type restrictions)
that are invoked in their proof.
To be more explicit and precise, we have the following algebraic relations
as well as the restrictions/EQM (which are exploited in the proof of the algebraic relations), namely;
\begin{eqnarray}
&&{\{s_b,s_{ad}}\}\,\bar C = 0\qquad\quad\;\Longleftrightarrow \qquad\quad{\cal B}\times C = 0,\nonumber\\
&&{\{s_b,s_{ad}}\}\bar\lambda = 0\qquad\quad\;\; \Longleftrightarrow\quad\qquad \partial_{\mu}D^{\mu}C = 0,\nonumber\\
&&[s_w,s_{ad}]\,A_{\mu} = 0\qquad\quad\Longleftrightarrow\qquad\quad {\cal B}\times C = 0,\nonumber\\
&&[s_w, s_{ad}]\,\bar\lambda = 0 \qquad\quad\;\; \Longleftrightarrow\quad\qquad
\partial_{\mu} D^{\mu} {\cal B}
 + \varepsilon^{\mu\nu}(\partial_{\nu}\bar C\times \partial_{\mu} C) = 0.
\end{eqnarray}
Thus, we observe  that the algebra (29) is very nicely respected  provided we utilize the 
strength of EQM from ${\cal L}_B^{(\bar\lambda)}$ and use the CF-type restrictions appropriately.

Now we focus on the ${\cal L}_{\bar B}^{(\lambda)}$ and 
briefly discuss the algebra of its symmetry operators. This Lagrangian density also respects
$five$ perfect symmetries. These are $s_d,s_{ad}$, 
$s_w = -\;{\{s_{ad},s_{ab}}\},$ $s_{ab}$ and\, $s_g$ [cf. Eqs. (2), (6), (27)]. In particular, we note that the anti-BRST symmetry transformations $(s_{ab})$ 
are same as (2) together with $s_{ab}\lambda = 0$
because we find that $s_{ab}({\cal B}\times \bar C) = 0$ due to the nilpotency
condition $s_{ab}^2{\cal B} = 0$.  The algebra satisfied by the above symmetry operators are:
\begin{eqnarray}
&&s_{(a)d}^2 =s_{ab}^2 = 0,\qquad\qquad{\{s_{ad},s_{ab}}\} = - s_w,\qquad {\{s_d,s_{ad}}\} = 0,\nonumber\\
&&[s_w,s_r] = 0,\qquad\quad r = d,ad,ab,g,\qquad\quad {\{s_d,s_{ab}}\} = 0,\nonumber\\
&&[s_g,s_d] = -s_d,\qquad[s_g,s_{ab}] = - s_{ab},\qquad\;\; [s_g,s_{ad}] = s_{ad}   \,.
\end{eqnarray}
We note that $s_{ad}\lambda=s_{ab}\lambda = 0$ implies that $s_w\lambda = 0$ because $s_w = -{\{s_{ab},s_{ad}}\}$.
We also have $s_g\lambda = +\lambda$ (i.e. the ghost number of $\lambda$ is +1).

From the above algebra, it is clear that we have found out
the  physical realizations of the cohomological operators
 $(d,\delta,\triangle)$ in the language of the symmetry transformations of the 
 Lagrangian density ${\cal L}_{\bar B}^{(\lambda)}$.
 However,  the algebra (32) is satisfied only when the EQM and 
the constraints (i.e. CF-type restrictions) of the theory are exploited together in a judicious manner.
We have been brief here in our statements but it can be 
easily checked that our claims are true. 
To be more explicit,
we note that we have obtained a {\it one-to-one} mapping: $d\longleftrightarrow  s_{ad}$\;, $\delta\longleftrightarrow s_{ab}$
and $\triangle\longleftrightarrow   s_w = -\;{\{s_{ab},s_{ad}}\}$. 
We conclude, from the above discussions, that the Lagrangian densities ${\cal L}_B^{(\bar\lambda)}$
and ${\cal L}_{\bar B}^{(\lambda)}$ respect $five$ perfect symmetries out of which 
$two$ are relevant fermionic symmetries (even though there is existence of three perfect fermionic symmetries present in the theory) 
and there is a unique bosonic symmetry $(s_w)$ in the theory. With these, we 
are able to provide the physical realization of the cohomological operators
$(d,\delta,\triangle)$. In other words, we have obtained $two$ independent
Lagrangian densities where the continuous symmetries provide the physical realizations
of the cohomological operators of differential geometry (at the algebraic level) which 
demonstrate that we have found out a 2D field theoretic model for the  Hodge theory (see, e.g. [5,7] for more details).

The identifications that have been made after equations (29) and (32) are correct in the language
 of continuous symmetries of the theory.
In this context, we have to recall our statements after Eq. (2) and Eq. (4)
 where we stated that the kinetic term and gauge-fixing term
remain invariant under the {\it fermionic} symmetries $s_{(a)b}$ and $s_{(a)d}$, respectively.
It is worth pointing out that the kinetic term owes its origin to the exterior derivative 
($d = dx^{\mu}, d^2 = 0$). On the other hand, the mathematical origin of the gauge-fixing term 
lies with the co-exterior derivative\footnote{The curvature 2-form 
$F^{(2)} = dA^{(1)}+ i A^{(1)}\wedge A^{(1)}$ (with $d = dx^{\mu}\partial_{\mu}$
 and $A^{(1)} = dx^{\mu}A_{\mu})$ leads to the derivation of the field strength tensor $F_{\mu\nu} =
 \partial_{\mu}A_{\nu}-\partial_{\nu}A^{\mu} + i\; ( A_{\mu}\times A_{\nu})$. Hence, the kinetic term
 owes its origin to $d = dx^{\mu}\partial_{\mu}$. It can be explicitly checked that $\delta A^{(1)} = -\star \; d  \;\star \;A^{(1)}  =
\partial_{\mu}A^{\mu}.$ Hence, the gauge-fixing term (i.e. a 0-form) has its origin in the co-exterior derivative $\delta  = -\star \; d \;\star.$ }
$(\delta  = - * d*, \delta ^2 = 0$.).
It is the ghost number considerations, at 
the level of charge, which leads to the identifications 
$ d\longleftrightarrow s_b,\quad\delta \longleftrightarrow s_d,\quad\triangle \longleftrightarrow s_w$ 
after the equation (29) as well as the mappings $ d\longleftrightarrow s_{ad},\quad\delta \longleftrightarrow s_{ab}
,\quad\triangle \longleftrightarrow s_w$ after the equation (32). 
Thus, the abstract mathematical cohomological operators find 
their realizations in the language of physically well-defined continuous 
symmetry operators of our present 2D non-Abelian 1-form gauge theory.

Now we concentrate on the algebraic structures associated with
the {\it six}  conserved charges (i.e. $Q_{(a)b},{Q_{(a)d}}, Q_w, Q_g$) that correspond to the
{\it six} continuous symmetries of our theory. We note that 
the nilpotency property of fermionic charges
$Q_{(a)b}$ and $Q_{(a)d}$ has already been quoted in Eq. (18). 
Using the expressions for the conserved
and nilpotent charges $Q_d$ and $Q_{ad}$ [cf. Eq. (17)] 
and the (anti-)co-BRST symmetry transformations (4), it can be 
readily checked that the  following is true as far as Lagrangian densities (1) are concerned, namely;
\begin{eqnarray}
&&s_{ad}\;Q_d = - i\;{\{Q_d,Q_{ad}}\} = 0, \qquad \Longrightarrow \qquad \mbox{iff} \qquad
\qquad \Longrightarrow \qquad {\cal B}\times C = 0,\nonumber\\
&&s_d \;Q_{ad} = - i\;{\{Q_{ad},Q_d}\} = 0, \qquad \Longrightarrow \qquad \mbox{iff} \qquad
\qquad \Longrightarrow \qquad {\cal B}\times \bar C = 0.
\end{eqnarray}
Thus, we note that even though the absolute anticommuting property $({\{s_d,s_{ad}}\}=0)$ associated with $s_{(a)d}$ 
is satisfied at the level of  symmetry operators {\it without} any use of CF-type 
restrictions $({\cal B}\times C = 0$, ${\cal B}\times\bar C = 0)$, we find that, at the level of conserved charges,
we have to exploit the strength of these restrictions (i.e. ${\cal B}\times C = 0$, ${\cal B}\times\bar C = 0$)
for the proof of absolute anticommutativity\footnote{The claims, made in Eq. (33), 
are {\it strong} statements. There are weaker versions of them which become
transparent when the operators $s_{(a)d}$ are applied on the {\it third} expressions for 
$Q_d$ and $Q_{ad}$ in (17). For instance, we note that $s_{ad}Q_d = i \int  dx\; \partial_1  [({\cal B}\times C)\cdot\bar C]
\longrightarrow 0$ for physicall well-defined fields that vanish off at $x = \pm \infty $.
Similarly, we observe that $s_dQ_{ad} = i\int dx \; \partial_1 [({\cal B}\times \bar C)\cdot C]\longrightarrow 0.$}.
 This is a {\it novel} observation which does {\it not} appear 
in the case of (anti-)BRST symmetries where ${\{s_b,s_{ab}}\} = 0$  and ${\{Q_b,Q_{ab}}\} = 0$
are satisfied 
{\it only} when the CF-condition $B+\bar B+( C\times \bar C) = 0$ is invoked. 
Another point to be noted is  that  the CF-type  restrictions ${\cal B}\times C = 0$ and ${\cal B}\times\bar C = 0$
are required for the proof of  $s_d\; Q_{ad} = -i {\{Q_{ad},Q_d}\} = 0$ and $s_{ad}\; Q_d = - i\;{\{Q_{ad},Q_d}\}$
as well as for the invariance of the Lagrangian densities  (i.e. $s_d \;{\cal L}_{\bar B}$ and $s_{ad}\,{\cal L}_ B$)
which are evident from Eq. (24).

The other algebraic relations amongst $Q_{(a)b}, Q_{(a)d}, Q_w$ and $Q_g$ are satisfied
in a straight-forward manner (except the absolute anticommutativity properties where the CF-type restrictions
are required). It can be checked that
\begin{eqnarray}
&&s_g Q_b =  - i\;[Q_b,Q_g] = + Q_b,\qquad\quad\quad s_g Q_{ad} = - i\,[Q_{ad},Q_g ] = +\; Q_{ad},\nonumber\\
&&s_g Q_{ab}= - i\;[Q_{ab},Q_g] = - Q_{ab},\quad\quad\quad s_g Q_d = - i\;[Q_d,Q_g] = - \;\;Q_d,\nonumber\\
&&s_g Q_w = -i\;[Q_w,Q_g] = 0,
\end{eqnarray}
which shows that the ghost number of $(Q_b,Q_{ad})$ is  equal to $( +1 )$
but the ghost number for $(Q_{ab},Q_d)$ is equal to $(-1)$ . It is also evident that $Q_w$
commutes with {\it all} the charges of the theory. 
As far as the proof of this statement is concerned, we note that
\begin{eqnarray}
&&s_w\;Q_r = - i\;[Q_r,Q_w] = 0,\qquad\qquad r = b,ab,d,ad,g,w,
\end{eqnarray}
which shows that $Q_w$ is the Casimir  operator for the whole algebra because it commutes with {\it all} the charges.
One of the simplest ways to prove this result is to compute the l.h.s. of equation (35) from the transformations 
(9) and the expressions for the charges 
$Q_r \;(r = b, ab, d , ad, g)$  that have been derived in Sec. 2.

We briefly comment here on the algebraic structure that is satisfied by the conserved charges of our theory.
 In this context, we have seen various forms of the algebras [cf. Eqs. (18), (34), (35)] that are satisfied by the 
{\it six} conserved charges of 
our theory. It can be verified that {\it collectively} these charges satisfy the following extended BRST algebra:
\begin{eqnarray}
&& Q_{(a)b}^2 = 0, \qquad Q_{(a)d}^2 = 0, \qquad \{ Q_b, Q_{ab} \} = \{ Q_d, Q_{ad} \} = 0, \nonumber\\
&& [Q_w, Q_r ] = 0, \qquad \qquad  r = b, ab, d, ad, g, w,   \quad \{ Q_d, Q_{ab} \} = 0, \nonumber\\
&& i\;[Q_g,Q_b] = +\; Q_b,\qquad\quad\quad  i\,[Q_{g},Q_{ad} ] = \; Q_{ad}, \quad \{ Q_b, Q_{ad} \} = 0, \nonumber\\
&&i\;[Q_{g},Q_{ab}] = -\; Q_{ab},\quad\quad\quad  i\;[Q_g ,Q_d] = - \;\;Q_d. 
\end{eqnarray}
The above algebra is obeyed {\it only} on a space of quantum fields defined in the 2D Minkowskian spacetime manifold where 
{\it all} types of CF-type restrictions as well as EOM, emerging from the Lagrangian densities (1), are satisfied.
The above algebra is  reminiscent of the Hodge algebra satisfied by the de Rham cohomological operators of 
differential geometry [18,19] where the mapping between the set of conserved charges and  cohomological operators is:
\begin{eqnarray}
(Q_b, Q_{ad}) \Leftrightarrow d, \quad  (Q_d, Q_{ab}) \Leftrightarrow \delta, \quad
Q_w = \{Q_b, Q_d\} = -\;\; \{Q_{ab}, Q_{ad}\} \Leftrightarrow \Delta.
\end{eqnarray}
This {\it two-to-one } mapping is true only for the coupled (but equivalent) 
Lagrangian densities (1) where the  EOM and CF-type restrictions are exploited together.

In the above identifications, the ghost number of a state 
(in the quantum Hilbert space), plays a very important 
role. We have shown in our earlier works [ 7, 20-22] that the algebra (36) 
indeed implies that if the ghost number
of a state $|\psi>_n$ is $n$ (i.e. $ i\; Q_g |\psi>_n = n \,  
|\psi>_n$), then, the states $Q_b |\psi>_n$, 
$Q_d |\psi>_n$ and $Q_w |\psi>_n$ would have the ghost numbers $(n + 1)$, $(n-1)$ and $n$, respectively.
In exactly similar fashion, we have already  been able to prove that the states $ Q_{ad} |\psi>_n$,
 $ Q_{ab}|\psi>_n$ and $Q_w|\psi>_n$ (with $Q_w = - {\{Q_{ab},Q_{ad}}\}$) would carry the ghost number
$(n + 1)$, $(n-1)$ and $n$, respectively\footnote {The above observations are the analogue of the operations of the 
de Rham
cohomological operators  $(d ,\delta ,\triangle )$ on the $n$-form $(f_n)$ where the
degrees of forms $df_n$, $\delta f_n$ and $\triangle f_n$ are $(n+1)$, $(n-1)$ and $n$, respectively.}. We have
discussed the Hodge decomposition theorem in the quantum 
Hilbert space of  states in our earlier works [7, 20-22]
which can be repeated for our 2D theory, too. This would fully 
establish the fact that our present theory is
a field theoretic model for the Hodge theory which 
provides the physical realizations of the cohomological 
operators in the language of symmetry transformations (treated as operators) and corresponding conserved charges.\\

\section {\bf Novel Observations:  Algebraic and  Symmetry Considerations in Our 2D non-Abelian Theory}

\noindent
As far as symmetry property is concerned, we observe that there are CF-type restrictions (${\cal B} \times C = 0,
 {\cal B} \times \bar C = 0$) corresponding to the {\it (anti-)co-BRST} symmetries, too, as is the case with the (anti-)BRST
symmetries of our 2D non-Abelian 1-form gauge theory where the CF-condition ($ B + \bar B + C \times \bar C = 0$)
exists [12]. However, there are specific novelties that are connected with the CF-type restrictions: ${\cal B} \times C = 0,
\;{\cal B} \times \bar C = 0$. First, these restrictions are (anti-)co-BRST invariant [i.e. 
$ s_{(a)d} ({\cal B} \times C) = 0$ and $ s_{(a)d} ({\cal B} \times \bar C) = 0$] whereas the
CF-condition ($ B + \bar B + C \times \bar C = 0$) is not {\it perfectly} invariant under 
the (anti-) BRST  transformations [cf. Eq. (25)]. Second, the restrictions
 (${\cal B} \times C = 0,\; {\cal B} \times \bar C = 0$) can be incorporated into the Lagrangian densities [cf. Eq. (26)] in such
a manner that one can have {\it perfect} (anti-)co-BRST symmetry invariance for the {\it individual} Lagrangian densities in (26).
Such kind of thing can {\it not} be done with the CF-condition $ B + \bar B +( C \times \bar C) = 0$.

We observe that (anti-)co-BRST symmetries (where the gauge-fixing term remains invariant) exist at the {\it quantum}
level when the gauge-fixing term is added to the Lagrangian densities. In other words, there is no {\it classical} 
analogue of the (anti-)co-BRST symmetries. However, the (anti-)BRST symmetry transformations (where the kinetic term
remains invariant) is the generalization  of the {\it classical} local $SU(N)$ gauge symmetries to the {\it quantum}
level. Furthermore, we note that the (anti-)BRST symmetries would exist for any $p$-form gauge theory in  {\it any}
arbitrary dimension of spacetime. However, the (anti-)co-BRST symmetries have been shown to exist  for
the $p$-form gauge theory {\it only} in  $D = 2p$ dimensions of spacetime [5,6]. They have {\it not} been shown to exist, so far,
in any {\it arbitrary} dimension of spacetime. In addition, the absolute anticommutativity property of the BRST and anti-BRST
 transformations require the validity of CF-condition. On the contrary, the nilpotent (anti-)co-BRST symmetries {\it do}
absolutely anticommute  with each other {\it without} any use of the CF-type restrictions that exist in the 2D non-Abelian gauge theory.

We note that ${\{s_d,s_{ad}}\}=  0$ without any use of the CF-type restrictions (${\cal B} \times C = 0,$ and
${\cal B} \times \bar C = 0$) as far as the Lagrangian densities ${\cal L}_B$ and ${\cal L}_{\bar B}$ [cf. Eq. (1)] 
 are concerned. However, the restrictions ${\cal B} \times C = 0$ and
${\cal B} \times \bar C = 0$ are required for the proof of ${\{Q_d,Q_{ad}}\} = 0$ when we compute this 
bracket from $s_d Q_{ad} = - i\;{\{Q_d,Q_{ad}}\}$ and/or $s_{ad} Q_d = -i\, {\{Q_{ad},Q_d}\}$
[as far as the Lagrangian densities ${\cal L}_B$ and ${\cal L}_{\bar B}$ [cf. Eq. (1)] are
concerned]. It is interesting to point out that the property of nilpotency and absolute anticommutativity is satisfied {\it without}
any use of CF-type restrictions for the Lagrangian densities (26) (where the Lagrange multipliers ${\lambda}$
and ${\bar\lambda}$ are incorporated to accommodate the CF-type restrictions). This statement is true for the
(anti-)co-BRST symmetry operators as well as for the corresponding conserved charges. The CF-type restrictions
$({\cal B}\times C = 0, {\cal B}\times C = 0)$ appear in the proof of ${\{Q_d, Q_{ad}}\} = 0$ [cf. Eq. (33)]
as well as in the mathematical expressions for $s_{ad}{\cal L}_B$ and $s_d {\cal L}_{\bar B}$ 
[cf. Eq. (24)] but they do {\it not} appear in ${\{s_d, s_{ad}}\} = 0$.
On the contrary, the CF-condition $(B + \bar B + C\times \bar C = 0)$ appears in the proofs of:${\{s_b, s_{ab}}\} = 0$,
 ${\{Q_b, Q_{ab}}\} = 0$ and in the mathematical expressions for: $s_{ab} {\cal L}_B$ as well as $s_b {\cal L}_{\bar B}$ [cf. Eq. (22)]
when the the Lagrangian densities (1) are considered.

To corroborate the above statements, we take a couple of examples to demonstrate that we do {\it not} require the strength 
of CF-type restrictions $({\cal B}\times C =0,\,  {\cal B}\times \bar C = 0 $ from outside) in the proof of nilpotency and absolute anticommutativity of the 
(anti-)co-BRST charges [derived from the Lagrangian densities (26)]. In this context, we note that the expressions for the nilpotent
 (anti-)co-BRST charges (17) remain the {\it same} for the Lagrangian densities (26) {\it but} the EOM [derived from 
(26)] are different from (12). We note that the latter are:
\begin{eqnarray}
&&\varepsilon^{\mu\nu} D_{\nu}{\cal B} +\partial^{\mu}B+(\partial^{\mu}\bar C\times C) = 0,
\quad \partial_{\mu}D^{\mu}C = 0,\quad {\cal B}\times C = 0,\nonumber\\
&& E = {\cal B} + (\bar{\lambda}\times C),\quad D_{\mu}\partial^{\mu}\bar C - i\,(\bar{\lambda}\times {\cal B}) = 0,\nonumber\\
&&\varepsilon^{\mu\nu}D_{\nu}{\cal B} -\partial^{\mu}\bar B - (\bar C\times\partial^{\mu} C) = 0,\quad \partial_{\mu}D^{\mu}\bar C = 0,\quad
{\cal B}\times \bar C = 0,\nonumber\\
&& E = {\cal B} + ({\lambda}\times\bar C),\qquad D_{\mu}\partial^{\mu}C + i\; ({\lambda}\times {\cal B}) = 0.
\end{eqnarray}
The above equations are to be used in the proof of conservation of the Noether currents from which the charges are computed.
In this context, we observe the expressions for the (anti-)co-BRST conserved Noether current for the Lagrangian density 
${\cal L}_B^{(\bar\lambda)}$ are as follows:
\begin{eqnarray}
&& J^{\mu(\bar\lambda)}_d = {\cal B}\cdot\partial^{\mu}\bar C -\varepsilon^{\mu\nu} B\cdot\partial_{\nu}\bar C
\equiv J^{\mu}_d \qquad  \mbox{ (cf.\; Eq. (16))},\nonumber\\
&&J_{ad}^{\mu(\bar\lambda)} = {\cal B}\cdot\partial^{\mu}C - \varepsilon^{\mu\nu}B\cdot\partial_{\nu}C
-\varepsilon^{\mu\nu}\bar C\cdot(\partial_{\nu}C\times C),
\end{eqnarray}
where the superscript ${(\bar\lambda)}$ denotes that the above currents have been derived
from ${\cal L}_B^{(\bar\lambda)}$ (cf. Eq. (26)). The  expressions (39) demonstrate that, for the Lagrangian density 
${\cal L}_B^{(\bar\lambda)}$ , the co-BRST Noether conserved current remains same as given in (16) (for ${\cal L}_B$)
{\it but} the expression for the anti-co-BRST Noether conserved current is {\it different}
from the {\it same} current derived from ${\cal L}^{(\bar\lambda)}_ B$ [cf. Eq. (16)].
The conservation of the above currents can be proven by using EL-EOM (38).
The expressions for the conserved co-BRST charge remains the {\it same} as given in (17) but the 
expression for the anti-co-BRST charge is: 
\begin{eqnarray}
Q_{ad}^{(\bar\lambda)} &=&\int dx\; J^{0(\bar\lambda)}_{ad}
\equiv \int dx \,\big[{\cal B}\cdot\dot C-\partial_1 B\cdot C + \bar C\cdot(\partial_1 C\times C)\big]\nonumber\\
&\equiv &\int dx\;\big[{\cal B}\cdot\dot C - D_0{\cal B}\cdot C +(\partial_1\bar C\times C)\cdot C +\bar C\cdot(\partial_1 C\times C)\big].
\end{eqnarray}
The nilpotency of the co-BRST charge $Q^{(\lambda)}_d = Q_d$ has already been proven in Eq. (18).
Similarly, it can be checked that
\begin{eqnarray}
s_{ad} \;Q_{ad}^{(\bar\lambda)} &=& s_{ad} \int\; dx \;[{\cal B}\cdot\dot C - D_0{\cal B}\cdot C +(\partial_1\bar C\times C)\cdot C +\bar C\cdot(\partial_1 C\times C)]\nonumber\\
&\equiv & \int\; dx\; \partial _1\; [i \;({\cal B}\times C)\cdot C]\longrightarrow 0\; \Longleftrightarrow \; -
 i\; {\{Q_{ad}^{(\bar\lambda)},Q_{ad}^{(\bar\lambda)}}\}=0,
\end{eqnarray}
which demonstrate the validity of nilpotency of $Q_{ad}^{(\bar\lambda)}$ because it can be explicitly
checked that $s_{ad} \;Q^{(\bar\lambda)}_{ad} = - i\; {\{Q^{(\bar\lambda)}_{ad}, Q^{(\bar\lambda)}_{ad}}\} = 0$
which implies that $(Q^{(\bar\lambda)}_{ad})^2 = 0$. We emphasize that the r.h.s. of (41) is zero due to the EOM
(i.e. ${\cal B}\times C = 0)$), too.

We now concentrate on the Lagrangian density ${\cal L}_{\bar B}^{(\lambda)}$ and compute the expressions
for the Noether currents corresponding to the (anti-)co-BRST symmetry transformations.
It is evident from transformations (27) that under the anti-co-BRST symmetry transformations, the Lagrangian density
 ${\cal L}_{\bar B}^{(\lambda)}$ transforms in exactly the same manner as given in (5).
Thus, the conserved current would be same as in (16). However, in view of the transformation of 
${\cal L}_{\bar B}^{(\lambda)}$ [in (28)] under $s_d$, we have the following expressions for the Noether current
\begin{eqnarray}
 J^{\mu(\lambda)}_d = {\cal B}\cdot\partial^{\mu}\bar C + \varepsilon^{\mu\nu}\bar B\cdot \partial_{\nu}\bar C
+\varepsilon^{\mu\nu}(\partial_{\nu}\bar C\times \bar C)\cdot C,
\end{eqnarray}
which is different from (16). The conservation law (i.e. $\partial_{\mu}J^{\mu(\lambda)}_d = 0$) can be proven 
by exploiting the EL-EOM given in (38). The conserved charge $Q^{(\lambda)}_d$ has the following forms:
\begin{eqnarray}
Q^{(\lambda)}_d &=&\int dx J^{0{(\lambda)}}_d
\equiv  \int\; dx \;\Big[{\cal B}\cdot \dot{ \bar C} + \partial_1 \bar B\cdot\bar C - (\partial_1\bar C\times \bar C)\cdot C\Big]\nonumber\\
&\equiv & \int\; dx \; \Big[{\cal B}\cdot\dot{\bar C} - D_0{\cal B}\cdot\bar C -(\bar C\times\partial_1 C)\cdot\bar C 
-(\partial_1 \bar C\times\bar C)\cdot C\Big],
\end{eqnarray}
where the EL-EOM have been used to obtain the above equivalent forms of the conserved charge . The nilpotency of the above charge can be proven by using the symmetry principle (with $s_d\;Q^{(\lambda)}_d =- i {\{Q^{(\lambda)}_d,Q^{(\lambda)}_d}\} = 0$) as:
\begin{eqnarray}
s_d\; Q^{(\lambda)}_d & = & s_d \;\int dx\;\big[{\cal B}\cdot\dot{\bar C} - D_0{\cal B}\cdot\bar C -(\bar C\times\partial_1 C)\cdot\bar C 
-(\partial_1 \bar C\times\bar C)\cdot C\big]\nonumber\\
&\equiv &\int dx\; \partial _1\; [i \;({\cal B}\times \bar C)\cdot\bar C]\longrightarrow 0.
\end{eqnarray}
Thus, we note that $s_d\; Q^{(\lambda)}_d = - i {\{Q^{(\lambda)}_d,Q^{(\lambda)}_d}\} = 0$ implies that
$(Q^{(\lambda)}_d)^2 = 0$. This proves the nilpotency of the co-BRST charge, derived from ${\cal L}_{\bar B}^{(\lambda)}$,
 for physically well-defined fields which vanish off  $x =\pm \infty$. Furthermore, the r.h.s. of (44) is zero due to the EOM
 (i.e. ${\cal B}\times C = 0$) which emerges from  ${\cal L}^{(\lambda)}_{\bar B}$ [cf. Eq. (38)].

We have to prove the absolute anticommutativity of the (anti-)co-BRST charges that 
have been derived from the Lagrangian densities
 (26). As pointed out earlier, the expressions for co-BRST charge for
 ${\cal L}^{(\bar\lambda)}_B$ remains the {\it same} as given in (17)
 (where there are primarily two equivalent expressions for it).
 We take, first of all, the following (with $Q^{(\bar\lambda)} = Q_d$) and apply the anti-co-BRST transformation $s_{ad}$:
 \begin{eqnarray}
s_{ad}\; Q^{(\bar\lambda)}_d &=& - i\; {\{Q^{(\bar\lambda)}_d,Q^{(\bar\lambda)}_d}\}
\equiv \int\; dx\;\big [ {\cal B}\cdot \dot {\bar C} - \partial_1 B\cdot \bar C\big]\nonumber\\
&\equiv & \int\; dx \big[{\cal B}\cdot (\dot{\cal B} -\partial_1 B)\big]. 
\end{eqnarray}
Using the equation of motion (38), the above expression yields
\begin{eqnarray}
&& s_{ad}\;Q^{(\bar\lambda)}_d =  i\,\int dx \big[({\cal B}\times C)\cdot\partial_1\bar C\big] = 0,
\end{eqnarray} 
due to the validity of EOM (i.e. ${\cal B}\times C = 0)$ w.r.t. $\bar\lambda$
from ${\cal L}^{(\bar\lambda)}_B$. Thus, we note that ${\{Q^{(\bar\lambda)}_d,Q^{(\bar\lambda)}_d}\} = 0$ on the 
{\it on-shell} for ${\cal L}^{(\bar\lambda)}_B$.
 In other words, the absolute anticommutativity is satisfied.
Now let us focus on the alternative expression for $Q^{(\bar\lambda)}_d$ and apply $s_{ad}$ on it:
\begin{eqnarray}
s_{ad}\; Q^{(\bar\lambda)}_d &=& \int\; dx\; \big[ {\cal B}\cdot\dot {\bar C} - D_0{\cal B}\cdot\bar C
+ (\partial_1\bar C\times C)\cdot\bar C\big]\nonumber\\
&\equiv & \int\; dx \;\partial_1 \big[({\cal B}\times C)\cdot{\bar C}\big] = 0.
\end{eqnarray}
Thus, we note that $s_{ad} \,Q^{(\bar\lambda)}_d =  - i\;{\{Q^{(\bar\lambda)}_d,Q^{(\bar\lambda)}_d}\} = 0$
for the physically well-defined fields that vanish off at $x =\pm \infty$. This absolute anticommutativity
is also  satisfied on the on-shell where ${\cal B}\times C = 0$ (due to the EOM from ${\cal L}^{(\bar\lambda)}_B$
w.r.t. $\bar\lambda$). Finally, we conclude that the property of absolute anticommutativity of the (anti-)co-BRST charges
 is satisfied {\it without }
invoking any  CF-type constraint condition from {\it outside}.

We now concentrate on the derivation of the absolute anticommutativity for $Q^{(\bar\lambda)}_d$
which is derived from ${\cal L}^{(\bar\lambda)}_B$. There are {\it two } equivalent expressions for it in Eq. (40).
We observe that the following are true, namely;
\begin{eqnarray}
s_d\; Q^{(\bar\lambda)}_d &=& \int \; dx \;s_d \big [ {\cal B}\cdot \dot C - \partial_1 B\cdot C
+\bar C\cdot (\partial_1 C\times C)\big]\nonumber\\
&\equiv & \int\; dx \,\partial_1\;\big [i \;\bar C\cdot ({\cal B}\times C)\big] = 0, 
\end{eqnarray}
where we have used the EOM from ${\cal L}^{(\bar\lambda)}_B$ w.r.t. $\bar\lambda$ that leads to 
${\cal B}\times C = 0$. Furthermore, for {\it all } the physically well-defined fields, we obtain 
$s_d\; Q_{ad}^{\bar\lambda}  = - i\; {\{Q_{ad}^{(\bar\lambda)} ,Q_d^{(\bar\lambda)}}\} = 0$
because {\it all} such fields vanish off at $x=\pm \infty$. Thus, the r.h.s. of  (48) is 
zero due to the Gauss's divergence theorem. Taking the alternative expressions for  
${Q_{ad}^{(\bar\lambda)}}$ in (40), we note that
\begin{eqnarray}
s_d \;Q^{(\bar\lambda)}_{ad} &=& \int\; dx s_d \;\big [ {\cal B}\cdot \dot C - D_0{\cal B}\cdot C + (\partial_1\bar C\times C)\cdot C
+ \bar C\cdot (\partial_1 C\times C)\big]\nonumber\\
&\equiv & \int \;dx \;\partial_1\; \big [- i\; ({\cal B}\times C)\cdot \bar C\big] = 0,
\end{eqnarray}
because of the fact that ${\cal B}\times C = 0$ (due to the EOM from ${\cal L}^{(\bar\lambda)}_B$
w.r.t. ${\bar\lambda}$ field).  Moreover, all the fields vanish-off at $x =\pm \infty $.
Thus, the Gauss divergence theorem shows that $s_d\;Q^{(\bar\lambda)}_{ad} =- i\; 
{\{Q_{ad}^{(\bar\lambda)} ,Q_d^{(\bar\lambda)}}\} = 0$ which proves the absolute anticommutativity property of 
the (anti-)co-BRST charges. This observation is a {\it novel} result in our present endeavor.

At this juncture, now we take up Lagrangian density ${\cal L}^{(\lambda)}_{\bar B}$ into consideration. The anti-co-BRST
charge for this Lagrangian density is same as given in (17) (i.e. $Q^{(\lambda)}_{ad} = Q_{ad}$).
We observe the following after the application of the co-BRST symmetry  $s_d$ on $Q_{ad}^{(\lambda)}$, namely;
\begin{eqnarray}
s_d\; Q^{(\lambda)}_{ad} &=&\int\; dx\; s_d\, \big[{\cal B}\cdot\dot C+ \partial_1\bar B\cdot C\big]\nonumber\\
&\equiv &\int dx \big [ i\; ({\cal B}\times\bar C)\cdot \partial_1 C \big ] = 0.
\end{eqnarray} 
Thus, we have seen now that $s_d\; Q^{(\lambda)}_{ad} \equiv - i\; {\{Q^{(\lambda)}_{ad},{Q^{(\lambda)}_d}\}} = 0$
due to ${\cal B}\times\bar C = 0$ which emerges as EOM from  ${\cal L}^{(\lambda)}_{\bar B}$ w.r.t. the field $\lambda$.
 In other words, the absolute anticommutativity ${\{Q^{(\lambda)}_{ad},{Q^{(\lambda)}_{ad}}\}} = 0$
 is satisfied on the {\it on-shell}. A similar exercise, with another equivalent expression for $Q^{(\lambda)}_{ad}$,
  namely;
\begin{eqnarray}
s_d\; Q^{(\lambda)}_{ad} &=& \int\; dx\; s_d\; \big [ {\cal B}\cdot\dot C - D_0{\cal B}\cdot C
- (\bar C\times \partial_1 C)\cdot C\big]\nonumber\\
&\equiv & \int\; dx \;\partial_1 \big [ i \,({\cal B}\times\bar C)\cdot C\big] = 0,
\end{eqnarray}
establishes the absolute anticommutativity  (i.e. ${\{Q^{(\lambda)}_{ad},{Q^{(\lambda)}_d}\}} = 0 $) due to Gauss's divergence theorem which states that {\it all } the physical fields vanish off at $x =\pm \infty$.

The absolute anticommutativity property can be {\it also} proven by using the expressions for the co-BRST charge 
$Q_d^{(\lambda)}$ [cf. Eq. (43)]. It can be readily checked that the following is true:
\begin{eqnarray}
s_{ad}\;Q_d^{(\lambda)} &=& \int\; dx\; s_{ad}\; \big [ {\cal B}\cdot \dot {\bar C} + \partial_1 \bar B\cdot\bar C
- (\partial_1\bar C\times\bar C)\cdot C\big]\nonumber\\
&\equiv & \int\; dx\; \partial_1 \big [ - i \,({\cal B}\times\bar C)\cdot C\big]\;\longrightarrow\;0. 
\end{eqnarray}
Thus, we note that $s_{ad}\;Q_d^{(\lambda)} = - i\;{\{Q_d^{(\lambda)},Q_{ad}^{(\lambda)}}\} = 0$
for the physically well-defined fields that vanish off at $x =\pm \infty$.
Moreover, the absolute anticommutativity is also satisfied due to EOM (i.e. ${\cal B}\times C = 0$)
that is derived from ${\cal L}^{(\lambda)}_{\bar B}$ w.r.t. Lagrange multiplier field  $\lambda$. Hence, the 
absolute anticommutativity of the (anti-)co-BRST charges is satisfied {\it on-shell}.
We now take up the alternative expression for the $Q_d^{(\lambda)}$ from (43)
and show the validity of absolute anticommutativity. Towards this goal in mind, we observe the following
\begin{eqnarray}
s_{ad}\; Q_d^{(\lambda)} &=&\int\; dx\; s_{ad}\; \big [  \;{\cal B}\cdot\dot {\bar C} - D_0{\cal B}\cdot\bar C
-(\bar C\times\partial_1 C)\times\bar C- (\partial_1\bar C\times\bar C)\cdot C\; \big]\nonumber\\
 &=&\int\; dx \;\partial_1\,\big [-i \;({\cal B}\times\bar C)\cdot C\big]\longrightarrow 0.
\end{eqnarray}
This shows that $s_{ad}\;Q_d^{(\lambda)} = - i\;{\{Q_d^{(\lambda)},Q_d^{(\lambda)}}\} = 0$
due to Gauss's divergence theorem which states that {\it all } the physical fields must vanish off at $x =\pm \infty $.
There is  another interpretation, too. The absolute anticommutativity (i.e. ${\{Q_d^{(\lambda)},\;Q_{ad}^{(\lambda)}}\}= 0)$
is satisfied {\it on-shell} (where ${\cal B}\times\bar C = 0$ due to EOM from ${\cal L}_{\bar B}^{(\lambda)}$ w.r.t. to $\lambda)$.

\section {\bf Conclusions}

\noindent
In our present endeavor, we have computed {\it all} the conserved charges of our theory and obtained the algebra 
followed by them. We have shown that, for the validity of the {\it precise } algebra (consistent with the algebra
obeyed by the cohomological operators), we have to use the 
EOM as well as the  CF-type restrictions of our theory described by the Lagrangian 
densities (1). In particular, we have demonstrated that the requirement
of the absolute anticommutativity property amongst the fermionic symmetry operators [cf. Eq. (31)] leads to the
emergence of our EOM and/or CF-type restrictions. In other words, it is the requirement of consistency of the operator algebra
with the Hodge algebra (i.e. the algebra obeyed by the  cohomological operators of differential geometry) that leads to the derivation
of the EOM as well as the CF-type restrictions of our theory. This way of derivation of the CF-type restrictions is completely different from our earlier derivations [10,11] where the existence of the continuous  symmetries (and their operator algebra) {\it and}
 the application of the superfield approach to BRST 
formalism have played key roles.

One of the highlights of our present investigation is the observation that the individual Lagrangian density 
[of the coupled Lagrangian densities  (26)] provides a model for the Hodge theory because the continuous symmetry
operators of the {\it specific} Lagrangian density (and corresponding charges) obey an algebra that is reminiscent of the 
algebra obeyed by the de Rham cohomological operators of differential geometry. In other words, the continuous symmetry
operators (and corresponding charges) provide the physical realizations of the cohomological operators of differential 
geometry. This happens because of the fact that the {\it individual} Lagrangian density respects {\it five }
perfect symmetries where there is no use of any kind of CF-type restrictions. This is precisely the reason
that {\it four} of the above mentioned {\it five } symmetries of the theory obey an 
$exact$ algebra that a reminiscent of the algebra obeyed by the de Rham cohomological operators of the differential geometry.

We have claimed in earlier works  [23,24] that the existence of the CF-restrictions is the hallmark of 
a {\it quantum } gauge theory (described within the framework of BRST formalism). This claim is as fundamental as the definition 
of a {\it classical } gauge theory in the  language of  first-class constraints by Dirac [25,26]. Thus, it has been a challenge for us to
derive {\it all} types of CF-type restrictions on our theory which respect the  (anti-)BRST as well as the (anti-)co-BRST symmetries {\it together}.
It is gratifying to state that we have discussed about the existence of CF-type restrictions from various points of view in
our works [10,11]. In fact, we have been able to show the existence of CF-type restrictions:  (i) from the symmetry considerations [10],
(ii) from the superfield approach to BRST formalism [11],  and (iii) from the algebraic considerations (in our present work). 
These works focus on  the importance of CF-type restrictions in the discussion of the 2D non-Abelian  theory.

As has been pointed  out earlier, one of the key features of (anti-)co-BRST symmetry transformations is the 
observation that these transformations absolutely anticommute {\it without} any use of CF-type restrictions ${\cal B}\times C = 0$
and ${\cal B}\times\bar  C = 0$. However, the latter appear very elegantly when we discuss the absolute anticommutativity 
of the co-BRST and anti-co-BRST charges in the language of symmetry transformations and their generators
[e.g. $s_d\,Q_{ad} = - i\,{\{Q_{ad},Q_d}\} = 0$ and $s_{ad}\,Q_d = - i\;{\{Q_d.Q_{ad}}\} = 0$].
This is a completely {\it novel} observation in our theory as it does {\it not} happen in the case of (anti-)BRST symmetry
transformations and in their absolute anticommutativity requirement. In fact, in the latter case of symmetries
[i.e. (anti-)BRST symmetries], 
the CF-condition is required for  the proof of the absolute anticommutativity of the (anti-)BRST charges
$\{ Q_b, Q_{ab} \} = 0 $ as well as  the (anti-)BRST symmetries $\{ s_b, s_{ab} \} = 0 $ (cf. Appendix A).

As far as the property of absolute anticommutativity and the existence of the 
CF-type conditions is concerned, we would like to
point out that the CF-type restrictions ${\cal B}\times C = 0$ and ${\cal B}\times\bar C = 0$ 
are invoked from 
{\it outside} in  the requirement of the absolute anticommutativity condition for the 
(anti-)co-BRST charges that are derived from the Lagrangian densities (1).
However, these restrictions are {\it not} required in the case of the absolute 
anticommutativity requirement of the (anti-)co-BRST charges that are derived from the Lagrangian densities (26).
This happens because of the observation that the CF-type restrictions:
 ${\cal B}\times C = 0$ and ${\cal B}\times\bar C = 0$ become 
EOM for the Lagrangian densities (26). All the tower of restrictions that have been derived
in [10,11] do {\it not} affect the d.o.f. counting for the gauge field because our present $2D$  non-Abelian 
gauge theory has been shown to be a {\it new} model of topological field theory where there are {\it no} propagating d.o.f. [7].  
Furthermore, the CF-type restrictions are amongst the auxiliary fields and (anti-)ghost fields which do 
{\it not} directly affect the d.o.f. counting for the gauge field of our theory.

We have been able to show the existence  of (anti-)BRST and (anti-)co-BRST symmetry transformations in the case of a 1D model of 
a rigid rotator [21]. However, the CF-type restriction, in the case of this 1D model
is {\it trivial} (as is the case with the Abelian 1-form gauge theory 
without any interaction with matter fields [7]). The non-trivial CF-type restrictions appear in the cases of 
6D Abelian $3$-form  and 4D  Abelian $2$-form gauge theories which have been shown to be the models for the Hodge 
theory within the framework of BRST formalism  [5,6].
It would be a nice future endeavor for us to apply our present ideas of 2D non-Abelian 1-form theory to the above
mentioned systems of physical interest. We are currently busy with these issues and our results would be reported in our future publications [27].
\\

\noindent
{\bf Acknowledgment}\\

\noindent
One of us (S. Kumar) is grateful to the BHU-fellowship under which the present investigation has been carried out.
The authors are thankful to N. Srinivas and T. Bhanja for fruitful discussions on the central theme of
the present research work  and thankfully acknowledge very useful comments and suggestions made 
by the Reviewer which have improved the quality of presentation of this paper.\\


\begin{center}
{\bf Appendix A: On  proof of $\{Q_b, \, Q_{ab}\} = 0 $}\\
\end{center}

\noindent
In this Appendix, we discuss a few essential theoretical steps to provide a proof for the absolute anticommutativity
of the conserved and nilpotent (anti-)BRST charges $Q_{(a)b}$. Towards this goal in mind,\,we observe
(with the input $s_b Q_{ab} = - i\;{\{Q_b,Q_{ab}}\}$) the following:
\[s_b \, Q_{ab} =\int \,dx \,s_b\,\Big[\dot{\bar B}\cdot\bar C -\bar B\cdot D_0\bar C
+\frac{1}{2}(\bar C\times \bar C)\cdot {\dot C}\Big].\eqno (A.1)\]
Using the BRST transformations from Eq.\,(2), we obtain the following explicit mathematical 
expressions from the\,{\it first} term (on the r.h.s. of the above equation):
\[s_b(\dot{\bar B}\cdot \bar C) = i\,(\dot{\bar B}\times C)\cdot \bar C
+ i ({\bar B} \times \dot C)\cdot \bar C +i\,\dot{\bar B}\cdot B.\eqno (A.2)\]
The \,{\it second} term, on the r.h.s of (A.1), leads to \hskip 0.5 cm
\[s_b\,(-\,\bar B\cdot D_0\bar C) = -\, i\,(\bar B\times C)\cdot {\dot{\bar C}} + 
(\bar B\times C)\cdot (A_0\times\bar C) - i\,{\bar B} \cdot {\dot B}~~~~~~~~~~~~~\]  
\[~~~~~~~~~~~~\equiv - i\,{\bar B} \cdot (\dot C\times\bar C) 
+\bar B\cdot[(A_0\times C)\times \bar  C+ \bar B\cdot(A_0\times  B),\eqno (A.3)\]
and the \,{\it third }term produces:
\[s_b\,\Big[\frac{1}{2}\,(\bar C\times \bar C)\cdot {\dot C}\Big] =  i\,(B\times\bar C)\cdot\dot C 
-\frac{i}{2}\,(\bar C\times\bar C)\cdot(\dot C\times C).\eqno(A.4)\]
Now we are in the position to apply the Jacobi identity to expand
\,$\bar B\cdot[(A_0\times C)\times \bar  C]$ 
and $\frac{i}{2}[(\bar C\times\bar C)\cdot(\dot C\times C)]$.
The outcome of these exercises  yield:
\[\bar B\cdot \Big[(A_0\times C)\times \bar C\Big] = -(A_0\times \bar B)\cdot (\bar C\times  C) - (A_0\times \bar C)\cdot (\bar B\times C),\]
\[-\frac{i}{2}\,\Big[(\bar C\times \bar C)\cdot\\
 (\partial_0 C \times C)\Big]= i\,(\partial_0 C\times \bar C)\cdot (\bar C\times \bar C).\eqno(A.5)\]

\noindent
The addition of {\it all} the terms with proper combinations, ultimately,\, leads to  the following explicit equation, namely;
\[i\,(\dot C\times\bar C)\cdot[B+\bar B +(C\times \bar C)]
- i\,\bar B\cdot D_0[B+(\bar C\times C)] + i\,\dot{\bar B}\cdot B
+i\,\dot{\bar B}\cdot(C\times\bar C).\eqno(A.6)\]
We note that the application of the CF-condition [i.e. $B+\bar B + (C\times \bar C) = 0$] produces the following:  
\[i\,\bar B\cdot\dot{\bar B}+ i\,\dot{\bar B}\cdot B-i\,\bar B\cdot\dot{\bar B}- i\,\dot{\bar B}\cdot B =0\equiv s_b \,Q_{ab},\eqno(A.7)\]
where we have used - $i \bar B\cdot D_0( B+C\times \bar C)= +\bar B\cdot D_0\bar B\equiv i\bar B\cdot\dot{\bar C}$.

In other words,\,we obtain  the relationship:
$s_b\,Q_{ab} = -i\;{\{Q_{ab},Q_b}\} = 0$\,\,(which is\,{\it true} only on the
submanifold of  fields in the quantum Hilbert space  where the CF-condition 
$B + {\bar B} + C \times {\bar C} = 0$ is 
satisfied). This is a reflection of the fact that the absolute anticommutativity of the 
(anti-)BRST transformations $\{s_b, s_{ab}\}\,A_{\mu} = 0$ 
is true only when the 
CF-condition $(B + {\bar B} + C \times {\bar C} = 0)$
 is imposed from {\it outside}. We conclude that the requirement of absolute anticommutativity condition for the nilpotent 
(anti-)BRST symmetry transformations is also reflected at the level of the requirement of the
absolute anticommutativity property of the off-shell nilpotent (anti-)BRST charges. The
CF-condition also appears in (22).

\begin{center}
{\bf Appendix B: On the derivation of}\, $ Q_w$\\
\end{center}
\noindent
In the main body of our text, we have derived the explicit expression for 
$Q_w$ from the Noether conserved current $(J_w)$. There is a simple way to obtain the same expression of 
$(Q_w)$  where the ideas behind the symmetry principle (and
concept of symmetry genrator) play an important role. In this context, we note the following:
\[s_d Q_b = -i\,{\{Q_b,Q_d}\} = -i\,Q_w,\quad\,s_b Q_d = -i\,{\{Q_d,Q_b}\} = - i\,Q_w.\eqno(B.1)\]
Thus, an explicit calculation of the l.h.s. [due to the transformations (2) and (4) as well as the expressions (13) and (17)]
yields the correct expression for $Q_w$. Let us, first of all, focus on the following:
\[s_b Q_d =\int dx\, s_b\,[{\cal B}\cdot\dot{\bar C} + B\cdot\partial_1\bar C].\eqno(B.2)\]
The {\it first} term produces the following explicit computations:
\[s_b({\cal B}\cdot\dot{\bar C}) = i\,({\cal B}\times C)\cdot\dot{\bar C} + i\,{\cal B}\cdot\partial_0B\]
\[= i\,({\cal B}\times C)\cdot\dot{\bar C} 
- i\,{\cal B}\cdot D_1 {\cal B}- i\,{\cal B}\cdot(\dot{\bar C} \times C) \equiv   -\,{\cal B}\cdot D_1 {\cal B},\eqno(B.3)\]
where we have used the EQM
\[\partial_0B = -D_1{\cal B} - (\dot{\bar C}\times C).\eqno(B.4)\]
The {\it second} term leads to 
\[s_b(i\,B\cdot\partial_1\bar C) = i B\cdot\partial_1 B.\eqno(B.5)\]
The addition of both the terms yield,
\[s_b\,Q_d = - i\,{\{Q_d,Q_b}\} = - i\,\int dx \Big[{\cal B}\cdot D_1{\cal B} - B\cdot\partial_1 B\Big],\eqno(B.6)\]
which, ultimately, leads  to the derivation of $Q_w$ [cf. Eq. (20)].

Now we dwell a bit on  the anticommutator  $s_d Q_b = - i\,{\{Q_b,Q_d}\} = -i\,Q_w$. 
In this connection,\,we have to use the symmetry transformations $(4)$ and expression $(13)$.
In other words, we compute the following:
\[s_d\, Q_b =\int dx\, s_d\,\Big[{\cal B}\cdot D_1 C + B\cdot D_0 C 
+\frac{1}{2}\,\dot{\bar C}\cdot(C\times C)\Big].\eqno(B.7)\]
 The {\it first} term, using the partial integration and dropping 
the total space derivative term, can be written in a different looking form
(i.e. ${\cal B}\cdot D_1 C = -D_1{\cal B}\cdot C)$. Now application of $s_d$
on the latter form, leads to the following explicit computation:
\[s_d \,(-D_1{\cal B}\cdot C) 
= i\,{\cal B}\cdot D_1{\cal B} 
+ i\,(\dot{\bar C}\times {\cal B})\cdot C.\eqno(B.8)\]
From the {\it second} and {\it third}  terms of (B.7), we obtain
\[s_d (B\cdot D_0 C)  = -\,{\cal B}\cdot \partial_0{\cal B} 
+ i\,B\cdot(\partial_1\bar C\times C) - B\cdot(A_0\times {\cal B}),\]
\[s_d\,\Big[\frac{1}{2}\,\dot{\bar C} \cdot(C\times C)\Big] =
i\,\dot{\bar C}\cdot({\cal B}\times C).\eqno(B.9)\]
Now, by using the equation of motion
\[D_0 {\cal B} =  \partial_1 B + (\partial_1\bar C\times C ),\eqno(B.10)\]
we observe  that  the sum of (B.8), (B.9) and (B.10) leads to the equality
 $[-i\,B\cdot \partial_1B]$. 
Thus, ultimately, we obtain  the following
\[s_d Q_b = -i\,{\{Q_b,Q_d}\} = -i\,Q_w, \]
where $(Q_w)$ [cf. Eq. (20)] is  $ Q_w = i\,\int dx \Big[{\cal B}\cdot D_1 {\cal B} - {\cal B}\cdot\partial_1 B\Big]$.
Thus, we have derived the precise form of $Q_w$ by using the ideas of continuous symmetries and their corresponding
generators. Thus, there are two distinct ways to derive $Q_w$.

\newpage
\begin{center}
{\bf Appendix C: On the (anti-)BRST symmetries of ${\cal L}^{(\bar\lambda)}_B$ and ${\cal L}^{(\lambda)}_{\bar B}$}
\end{center}
\noindent
We have observed earlier that the coupled Lagrangian densities (26) respect {\it five}
perfect symmetries {\it individually}. As far as the (anti-)BRST symmetries are concerned, we
have noted that ${\cal L}^{(\bar\lambda)}_B$ respects {\it perfect} BRST symmetries, listed
in (2), along with $s_b \;\bar\lambda =  0$\footnote {Because it transform to a total spacetime derivative (i.e. $s_b {\cal L}^{(\bar\lambda)}_B
= \partial_{\mu}[B\cdot D^{\mu}C]$).}. We discuss here the anti-BRST symmetry of {\it this} Lagrangian density.
 It can be checked  that, under the anti-BRST symmetry transformations (2)
along with $s_{ab}\;\bar\lambda = - i \;(\bar\lambda\times\bar C)$, we have the following transformation for 
the Lagrangian density ${\cal L}^{(\bar\lambda)}_B$:
\[s_{ab}\;{\cal L}^{(\bar\lambda)}_B = \partial_{\mu} \big [ -(\bar B + C\times\bar C)\cdot\partial^{\mu}\bar C\big] 
 +(B+\bar B+C\times\bar C)\cdot D_{\mu}\partial^{\mu}\bar C\]
\[ - i\;\bar\lambda
\cdot({\cal B}\times {\{\bar B+(C\times\bar C)}\}).\eqno(C.1)\]
If we implement the CF-condition $B+\bar B+(C\times\bar C)=0$, we obtain the following,
from the above  transformation  of ${\cal L}^{(\bar\lambda)}_B$, namely;
\[s_{ab}\;{\cal L}^{(\bar\lambda)}_B =\partial_{\mu} \big [B\cdot\partial^{\mu}\bar C\big] +
i\;\bar\lambda\cdot({\cal B}\times B).\eqno (C.2)\]
For the anti-BRST invariance, we impose a new CF-type restrictions [i.e. $\bar\lambda\cdot({\cal B}\times B) = 0$]
which involves {\it three} auxiliary fields. As a consequence, this restriction can be equivalent to the 
following {\it three} individual constraints in terms of {\it only} {\it two} auxiliary fields, namely;
\[\bar\lambda\cdot({\cal B}\times B) = 0\quad\Longrightarrow\quad {\cal B}\times B = 0,\qquad 
\lambda\times B = 0,\qquad \lambda\times {\cal B}= 0.\eqno (C.3)\] 
The above restrictions have been derived from the symmetry point of view [10] as well as by using the 
augmented version of  superfield formalism [11]. It is gratifying to note that the 
(anti-)BRST symmetry transformations (that include transformations on $\lambda$ and $\bar\lambda$) absolutely
anticommute if we consider the above restrictions.

Now we focus on the Lagrangian density ${\cal L}^{(\lambda)}_{\bar B}$.
 It has perfect anti-BRST invariance with $s_{ab} \lambda = 0$ [and $s_{ab} ({\cal B}\times \bar C) = 0$].
 We discuss here the application of BRST transformation (2), along with $s_b\;\lambda = - i\;(\lambda\times C)$,
 on ${\cal L}^{(\lambda)}_{\bar B}$. This exercise leads to the following;
 \[s_b \,{\cal L}^{(\lambda)}_{\bar B}= \partial_{\mu}\big[(B+C\times \bar C)\cdot\partial^{\mu}C\big]
 -(B+\bar B+C\times\bar C).D_{\mu}\partial^{\mu}C\]
 \[-i\lambda\cdot\big[{\cal B}\times{\{B+(C\times\bar C)}\}\big].\eqno (C.4)\]
If we impose the usual CF-condition $B+\bar B+(C\times\bar C) = 0$ from outside on (C.4), we obtain the following from the 
above transformation of ${\cal L}^{(\lambda)}_{\bar B}$, namely;
\[s_b \;{\cal L}^{(\lambda)}_{\bar B} = \partial_{\mu}\big[-\bar B\cdot\partial^{\mu} C\big]
+i\lambda\cdot({\cal B}\times\bar B).\eqno (C.5)\]
Thus, for the BRST invariance of the action integral $S =\int d^2x \;{\cal L}^{(\lambda)}_{\bar B}$,
 we invoke another CF-type restriction:
\[\lambda\cdot({\cal B}\times\bar B) = 0
\qquad \Longrightarrow \qquad {\cal B}\times \bar B = 0,\quad
\lambda\times {\cal B} = 0,\quad \lambda\times\bar B = 0.\eqno (C.6)\]
In the above, we have noted that there are two constraint restrictions 
(i.e. $B+\bar B+C\times \bar C$,\quad $\lambda\cdot({\cal B}\times\bar B) = 0$) that ought to be invoked
for the BRST invariance of the action integral. It is clear that the latter restriction involves 
{\it three} auxiliary fields. However, this restriction {\it actually} corresponds to {\it three}
CF-type restrictions that have been written in (C.6). The latter {\it three} CF-type  restrictions
are correct as they have been derived from the symmetry consideration in [10]. It is gratifying to state, at
this juncture, that the restrictions, listed in (C.3) and (C.6), are required for the absolute anticommutativity
of the (anti-)BRST symmetries (2) along with $s_b\;\lambda = - i\;(\lambda\times C)$ and 
$s_{ab}\;\bar\lambda = -i\,(\bar\lambda\times\bar C)$.


\begin{thebibliography}{99}
\bibitem{sun1}   C. Becchi, A. Rouet, R. Stora,  {\it Phys. Lett. }  B {\bf 32}, 344 (1974).
\bibitem{sun2}   C. Becchi, A. Rouet,  R. Stora,  {\it Commun. Math. Phys.}  {\bf 42}, 127 (1975).
\bibitem{sun3}   C. Becchi, A. Rouet, R. Stora,  {\it Ann. Phys. (N. Y.)}  {\bf 98}, 287 (1976).
\bibitem {sun4}  I. V. Tyutin, Lebedev Institute Report, Preprint FIAN-39, 1975 (Unpublished).
\bibitem{DJ}     R. P. Malik,  {\it Int. J. Mod. Phys.} A  {\bf 22}, 3521 (2007).
\bibitem{gts}    R. Kumar, S. Krishna, A. Shukla, R. P. Malik,\\
                   {\it Int. J. Mod. Phys.} A  {\bf 29}, 1450135 (2014).
\bibitem{tfc}    R. P. Malik,  {\it J. Phys. A: Math. Gen.}  {\bf 34}, 4167 (2001).
\bibitem{sun1}   E. Witten,  {\it Nucl. Phys.} B  {\bf 202}, 253 (1982).
\bibitem {sun2}  A. S. Schwarz,  {\it Lett. Math. Phys.}  {\bf 2}, 217 (1978).             
\bibitem{ftc1}   N. Srinivas, S. Kumar, B. K. Kureel, R. P. Malik, \\{\it Int. J. Mod. Phys.} A {\bf 32}, 1750136  (2017) A 32: 1750193.
\bibitem{sun3}   N. Srinivas, R. P. Malik,{\it Int. J. Mod. Phys.} A {\bf 32}, 1750193 (2017). 
\bibitem{sun4}   G. Curci, R. Ferrari,  {\it Phys. Lett.} B  {\bf 63}, 91 (1976). 
\bibitem{sun 5}  N. Nakanishi, I. Ojima,  {\it Covariant Operator Formalism\\ of Gauge
                 Theories and Quantum Gravity (World Scientific, Singapore, 1990)}. 
\bibitem{sun6}   K. Nishijima,  {\it Czech. J. Phys.}  {\bf 46}, 1 (1996).
\bibitem{sun 7}  D. Dudal, V. E. R. Lemes, M. S. Sarandy, \\S. P. Sorella, M. Picariello,
                  {\it JHEP}  {\it 0212}, 008  (2002).
\bibitem{sun 8}  D. Dudal, H. Verschelde, V. E. R. Lemes, M. S. Sarandy, S. P. Sorella, M. Picariello,\\
                 A. Vicini, J. A. Gracey,  {\it JHEP}  {\bf 0306}, 003 (2003).
\bibitem{sun 9}  R. Kumar, S. Gupta, R. P. Malik,  {\it Int. J. Theor. Phys.}    {\bf 55}, 2857 (2016).
\bibitem{sun 31} T. Eguchi, P. B. Gilkey, A. Hanson,  {\it Phys. Rep.}  {\bf 66}, 213 (1980).
\bibitem{sun32}   S. Mukhi, N. Mukunda,  {\it Introduction to Topology, Differential Geometry and Group
                 Theory for Physicists (Wiley Eastern Private Limited, New Delhi, 1990)}.
\bibitem{sun 10} S. Gupta, R. P. Malik,  {\it Eur. Phys. J.} C  {\bf 58}, 517 (2008).
\bibitem{sun11}  R. P. Malik,  {\it Int. J. Mod. Phys.} A  {\bf 5}, 1685 (1998).
\bibitem{sun 12} S. Gupta, R. P. Malik,  {\it Eur. Phys. J.} C  {\bf 68}, 325 (2010).
\bibitem{sun 13} L. Bonora, R. P. Malik,  {\it Phys. Lett.} B  {\it 655}, 75 (2007).
\bibitem{sun 14} L. Bonora, R. P. Malik,  {\it J. Phys. A: Math. Theor.} {\bf 43},  375 (2010). 
\bibitem{sun 15} P. A. M. Dirac,  {\it Lectures on Quantum Mechanics, Belfer
                 Graduate School of Science (Yeshiva University Press, New York, 1964)}.
\bibitem{sun 16}  K. Sundermeyer,  {\it Constrained Dynamics: Lecture Notes
                  in Physics, Vol.  169\\  (Springer, Berlin, 1982)}.
\bibitem{sun 17}  R. P. Malik,  etal., in preperation.                  
\end{thebibliography}
\end{document}